%% file: main.tex
\documentclass[sigconf,balance=false, nonacm]{acmart}
\usepackage{popets}

\input{preamble}

\usepackage{xspace}
\newcommand{\IW}{\texttt{TRIDENT}\xspace}
\setcopyright{popets}
\copyrightyear{YYYY}

\acmYear{YYYY}
\acmVolume{2025}
\acmNumber{4}
\acmDOI{XXXXXXXXXXXXXX}
\acmISBN{}
\acmConference{Proceedings on Privacy Enhancing Technologies}
\begin{document}

\title{TRIDENT: Tri-modal Real-time Intrusion Detection Engine for New Targets}


\author{Ildi Alla\textsuperscript{*}, Selma Yahia\textsuperscript{*}, Valeria Loscri\textsuperscript{*}}

\affiliation{%
  \institution{\textsuperscript{*}Inria Lille-Nord Europe, Lille, France}
  \country{}
}
\email{{ildi.alla, selma.yahia, valeria.loscri}@inria.fr}


\renewcommand{\shortauthors}{Alla et al.}

\begin{abstract}

The increasing availability of drones and their potential for malicious activities pose significant privacy and security risks, necessitating fast and reliable detection in real-world environments.
However, existing drone detection systems often struggle in real-world settings due to environmental noise and sensor limitations. 
This paper introduces \IW, a tri-modal drone detection framework that integrates synchronized audio, visual, and RF data to enhance robustness and reduce dependence on individual sensors. \IW introduces two fusion strategies—Late Fusion and GMU Fusion—to improve multi-modal integration while maintaining efficiency. 
The framework incorporates domain-specific feature extraction techniques alongside a specialized data augmentation pipeline that simulates real-world sensor degradation to improve generalization capabilities. 
A diverse multi-sensor dataset is collected in urban and non-urban environments under varying lighting conditions, ensuring comprehensive evaluation. Experimental results show that \IW achieves 96.89\% accuracy in real-world recordings and 83.26\% in a more complex setting (augmented data), outperforming unimodal and dual-modal baselines. Moreover, \IW operates in real-time, detecting drones in just 6.09 ms while consuming only 75.27 mJ per detection, making it highly efficient for resource-constrained devices. 
The dataset and code have been released to ensure reproducibility\footnotemark.

\footnotetext{\url{https://github.com/TRIDENT-2025/TRIDENT}}

\end{abstract}

\maketitle

\vspace{-0.2cm}
\section{Introduction and Motivation}
The proliferation of Unmanned Aerial Vehicles (UAVs), commonly known as drones, has introduced transformative capabilities across various sectors, including package delivery, infrastructure inspection, and environmental monitoring \cite{emimi2023current, GARG2023103831}. This rapid advancement has also intensified serious privacy vulnerabilities \cite{pillai2024privadome, tedeschi2024selective}. Modern drones, equipped with high-resolution cameras and advanced sensors, have been involved in unauthorized surveillance. Reports have documented drones hovering near residential areas in the UK, capturing private footage of residents and celebrities without consent \cite{first-uk,bbc-drone}. Similar intrusions have been reported in U.S. neighborhoods, where drones have disrupted private gatherings, raising concerns over covert surveillance and data theft \cite{cisa-drone-2024,airsight-drones-security-2018}. Beyond privacy issues, drones have also been weaponized, with their unauthorized use in restricted airspace posing significant threats to national security \cite{cameron-intel-officials-2024,swope-unexplained-drones-2024}. Their ability to operate at low altitudes, maneuver discreetly, and evade conventional security systems highlights the urgent need for advanced detection technologies to reliably identify and mitigate these evolving threats.

Current drone detection systems generally fall into single-sensor and multi-sensor approaches. Single-sensor methods, including acoustic, visual, and Radio Frequency (RF)-based techniques, are simple to deploy but highly sensitive to environmental interference. 
Acoustic systems struggle in noisy urban settings, leading to false detections or missed targets \cite{shi2018hidden,anwar2019machine}. Similarly, visual-based methods, such as Convolutional Neural Networks (CNNs),
perform well in controlled conditions but become unreliable in low-light or occluded settings \cite{huang2024all, mrabet2024machine,misbah2024msf}. RF-based systems, which analyze UAV communication signals, are prone to interference in congested areas, where overlapping signals can mask drone transmissions \cite{ezuma2019detection,medaiyese2021semi}.  Multi-sensor systems aim to overcome these limitations by integrating multiple modalities. Dual-modality approaches, such as audio-visual or RF-audio fusion, improve detection by combining complementary information \cite{lee2023cnn, frid2024drones}. However, they remain unreliable in challenging environments where noise, obstructions, or overlapping RF signals compromise both sensing modalities \cite{jovanoska2021passive}. Tri-modality systems
enhance detection by increasing robustness 
and capturing a richer set of information \cite{svanstrom2021real, jovanoska2021passive, mccoy2024optimized}.

Despite their advantages, current tri-modal approaches fall short in addressing the complexities of real-world drone detection due to three primary challenges.
\textbf{First}, the datasets used to train and evaluate these systems often do not adequately represent the complexity of the real world. Many are collected in controlled settings—such as open fields or designated test areas—with minimal environmental variability \cite{svanstrom2021real, shi2018anti}. These conditions do not reflect the diverse noise profiles, visual clutter, and RF interference typical of urban and semi-urban environments \cite{mccoy2024optimized}. Furthermore, these datasets generally lack synchronized multi-modal degradation, where disruptions across different sensing modalities occur simultaneously. This limitation hinders model generalization, reducing reliability in unpredictable operational deployments.
\textbf{Second}, prevailing fusion techniques employed in tri-modal systems are often insufficiently sophisticated for robust multi-modal integration. Existing systems often rely on basic fusion strategies, such as simple averaging or concatenation of sensor output at the decision level \cite{svanstrom2021real, shi2018anti}. These rudimentary techniques fail to fully leverage the rich, complementary information inherent in heterogeneous sensor data and lack the capacity for deep cross-modal feature interaction and dynamic adaptation to varying sensor reliability under different environmental conditions. \textbf{Third}, comprehensive data augmentation strategies, crucial to improving model robustness and generalization, are conspicuously absent in existing tri-modal drone detection research \cite{lee2023cnn, ding2023drone}. Real-world drone detection systems must operate reliably despite significant environmental variations, including fluctuating noise levels, variable lighting conditions, occlusions from urban structures, and diverse RF interference. The lack of comprehensive data augmentation in current methodologies severely limits their ability to adapt to these unseen, challenging environments.

This paper 
 introduces \IW, an effective tri-modal UAV detection framework designed to overcome the limitations of existing systems and enable robust, real-time drone detection in diverse and complex environments. \IW addresses the above critical challenges through a comprehensive approach that integrates novel data acquisition, enhanced pre-processing and augmentation techniques, and sophisticated fusion techniques, ensuring greater adaptability and reliability in real-world scenarios.

\smallskip
\noindent \textbf{Ethics:}~All experiments were conducted under the necessary permissions
and do not raise any ethical concerns.
\smallskip

\noindent \textbf{Summary of Contributions} 
\smallskip



\noindent $\bullet$ We propose \IW, a novel framework for real-time drone detection designed for practical deployment. \IW ensures robust detection by integrating data from audio, visual, and RF sensors, implementing two distinct fusion strategies: Late Fusion, which efficiently combines final predictions from unimodal models, and GMU Fusion, which integrates intermediate-layer features.
Both fusion approaches are optimized for energy-efficient model architectures, making \IW well-suited for privacy-preserving security applications with limited computational resources.


\noindent $\bullet$
We introduce an audio feature extraction scheme, a specialized pre-processing method for RF signals, and a frame-by-frame analysis approach for visual data. Further, we design a targeted data augmentation pipeline that applies synchronized, modality-consistent noise levels—both low and high—across inputs within each sample. This pipeline provides worst-case real-world degradation and measures the lower bound \IW accuracy. Crucially, these augmentations are used \emph{exclusively at test time}, while all models are trained on real, unaugmented data.

\noindent $\bullet$ 
We collect a 10 GB multi-sensor dataset, a key enabler for advancing research in robust drone detection. This dataset is uniquely distinguished by its diverse real-world collection settings, spanning both complex urban and less-obstructed non-urban locations.
Furthermore, data acquisition was conducted under varied lighting conditions, including daylight and sunset scenarios, to capture the dynamic challenges of real-world deployments. 
\noindent $\bullet$ Our experimental results demonstrate the strong performance advantages of \IW. Specifically, our Late Fusion approach achieves an accuracy of 96.89\% for tri-modal fusion in real-world data
and 83.26\% in high noisy augmented data.
Beyond its high accuracy, this approach operates in real time, requiring only 6.09 ms per detection in its optimized configuration. Moreover, energy consumption analysis on a resource-constrained device highlights its efficiency, consuming as little as 75.27 mJ per detection.

\vspace{-0.2cm}
\section{Background on Drone Detection}
\noindent \textbf{Technical Challenges.}
As discussed in Section \ref{Related Work}, existing drone detection techniques struggle in real-world environments, where model generalization issues, real-time processing constraints, and sensor-specific degradations impact detection accuracy and efficiency. Our work directly addresses the following key technical challenges inherent in real-world UAV detection:

\noindent $\bullet$ \textbf{Environmental Noise and Interference.}~
Drone detection models must operate reliably in \textit{noisy} and \textit{complex environments}, where multiple factors degrade sensor performance. Acoustic-based detection is affected by traffic, human speech, wind, and wildlife, which can overlap with drone sounds, leading to false negatives and reduced precision \cite{casabianca2021acoustic, ohlenbusch2021robust}. Visual-based detection struggles with occlusions from buildings and trees, as well as lighting variations like shadows, glare, and low light, which reduce accuracy \cite{amala2023drone, liu2024vision}. RF-based detection, which analyzes drone communication patterns, faces significant interference in the crowded 2.4 GHz ISM band, where Wi-Fi, Bluetooth, and other wireless devices operate \cite{nguyen2025effective,gluge2024robust}. This high-density RF environment complicates the reliable isolation of drone signals from background noise, further challenging detection accuracy.

\noindent $\bullet$ \textbf{Model Generalization.}~ 
Machine learning models for drone detection often struggle to generalize when deployed in unfamiliar environments that differ from their training data. Data augmentation, such as adding synthetic noise, is commonly used to improve generalization but often fails to fully bridge the gap between idealized training conditions and real-world complexities  \cite{wisniewski2024drone}.
One major challenge is \textit{overfitting to synthetic noise}. Models trained on artificially generated noise often learn its specific characteristics, making them highly sensitive to these distortions while failing to adapt to the diverse and unpredictable noise variations encountered in real-world settings. This rigidity in learned patterns results in a significant performance drop when exposed to unseen noise conditions during deployment. Another limitation is the \textit{incomplete representation of real-world degradations}. Many augmentation techniques modify only one sensor modality at a time, overlooking the synchronized degradations that impact multiple sensors simultaneously. For example, in urban environments, visual occlusions from buildings or trees often coincide with acoustic masking from traffic or construction noise.

\noindent $\bullet$ \textbf{Real-time Processing Constraints.}~ 
Effective drone detection for security and privacy applications demands \textit{real-time} or \textit{near real-time performance}, ensuring timely responses to unauthorized drone activity. However, achieving this speed is challenging due to two primary factors. 
First, deep learning models, while offering high detection accuracy, require substantial computational power, often creating bottlenecks that lead to delays—particularly on resource-constrained edge devices.
Second, multi-sensor data synchronization and fusion latency further hinder real-time performance. Integrating data from heterogeneous sensors in real-time requires precise temporal alignment and efficient fusion mechanisms. Without effective synchronization, delays in processing can degrade the system’s responsiveness, 
limiting its reliability in operational scenarios.

\vspace{-0.35cm}
\section{Threat Model}\label{Threat Model}
The increasing accessibility and sophistication of drone technology have created a tangible and evolving threat landscape for both individual privacy and organizational security. This work considers a realistic threat model involving a ground-based drone detection station safeguarding a defined area from unauthorized UAV activities.  We posit an adversary seeking to exploit drone technology for malicious purposes, characterized by the following realistic knowledge and capabilities, denoted as \( \mathcal{K} = \{I, L, C, T, S\} \):

\noindent $\bullet$ \textbf{ \( I \) - Drone Identification:}  The adversary knows specific drone \textit{types} and \textit{models} commonly used for illicit activities, as these are often commercially available and easily accessible.


\noindent $\bullet$ \textbf{ \( L \) - Target Locations:}  
 The adversary identifies specific \textit{target locations} for surveillance or intrusion, such as private properties, restricted airspace, or sensitive facilities, using publicly available maps and aerial imagery.


\noindent $\bullet$ \textbf{ \( C \) - Drone Capabilities:}
The adversary has a baseline understanding of common drone \textit{capabilities}, such as flight range, endurance, payload capacity, and sensor specifications, which are easily accessible from public technical documentation.


\noindent $\bullet$ \textbf{ \( T \) - Flight Trajectories:}  The adversary can plan sophisticated \textit{flight trajectories}, including low-altitude paths to evade radar detection and autonomous navigation to reduce RF communication. Knowledge of flight patterns and restricted zone entry/exit points can be derived from public airspace information and observations.


\noindent $\bullet$ \textbf{ \( S \) - Communication Signals:}   The adversary knows the typical \textit{operating frequency bands} for drones, such as the 2.4 GHz and 5.8 GHz ISM bands, identifiable through publicly available information.

With this knowledge, \( \mathcal{K}\), adversaries can engage in privacy-invasive activities, such as collecting private images and videos, intercepting communications, or exploiting flight paths to infiltrate restricted zones. This threat model underscores the need for robust detection systems to protect individual privacy and organizational security.
\vspace{-0.2cm} 
\section{Dataset Collection}
\label{measurement setup}
\subsection{Measurement Locations}
\begin{figure}[h]
    \centering
    \vspace{-0.25cm} 
    \includegraphics[scale=0.3]{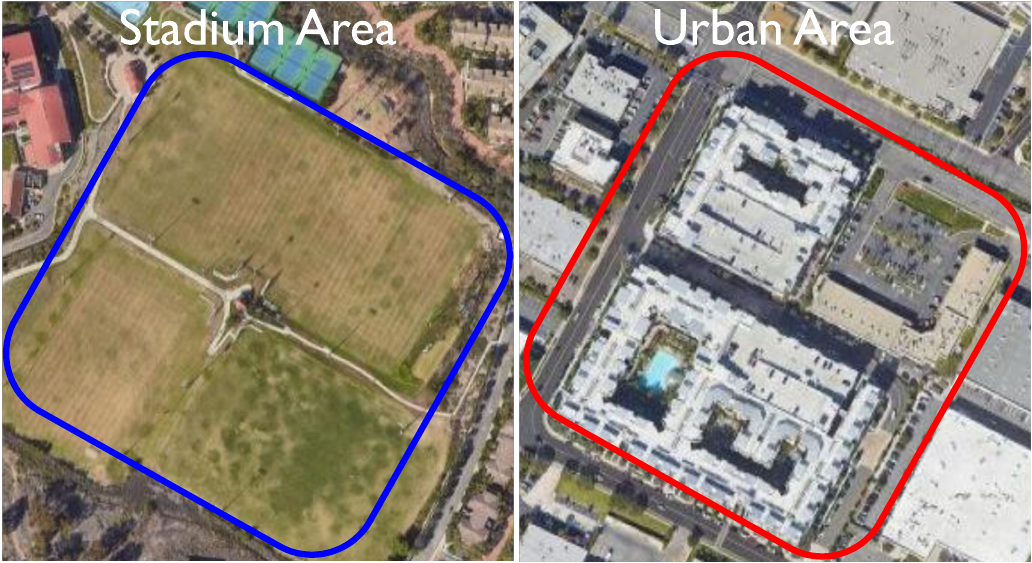}
    \vspace{-0.4cm} 
    \caption{Measurement locations during data collection.}
    \label{fig:measurement_setup_locations}
    \vspace{-0.3cm} 
\end{figure}

Data collection was conducted in two primary locations: a non-urban area (stadium) and an urban area, as illustrated in Figure \ref{fig:measurement_setup_locations}. These locations were carefully selected to account for varying environmental complexities, noise levels, and potential interference. 

\noindent \textbf{Non-Urban Area.} Data was gathered in a stadium surrounded by trees and grass, offering a controlled environment with minimal urban interference. Background noise was moderate, primarily from nearby roads and occasional human or wildlife activity, making it ideal for baseline measurements.


\noindent \textbf{Urban Area.} Measurements were conducted 3.2 km from an airport in a busy urban setting with high noise levels from aircraft, traffic, and electronic devices. This environment introduced challenges such as overlapping audio frequencies, visual distractions, and RF interference, evaluating the system’s performance in complex real-world conditions.




\vspace{-0.2cm} 
\subsection{Drones and Sensing Equipment}

\begin{figure}[h]
    \centering
    \vspace{-0.25cm} 
    \includegraphics[scale=0.2]{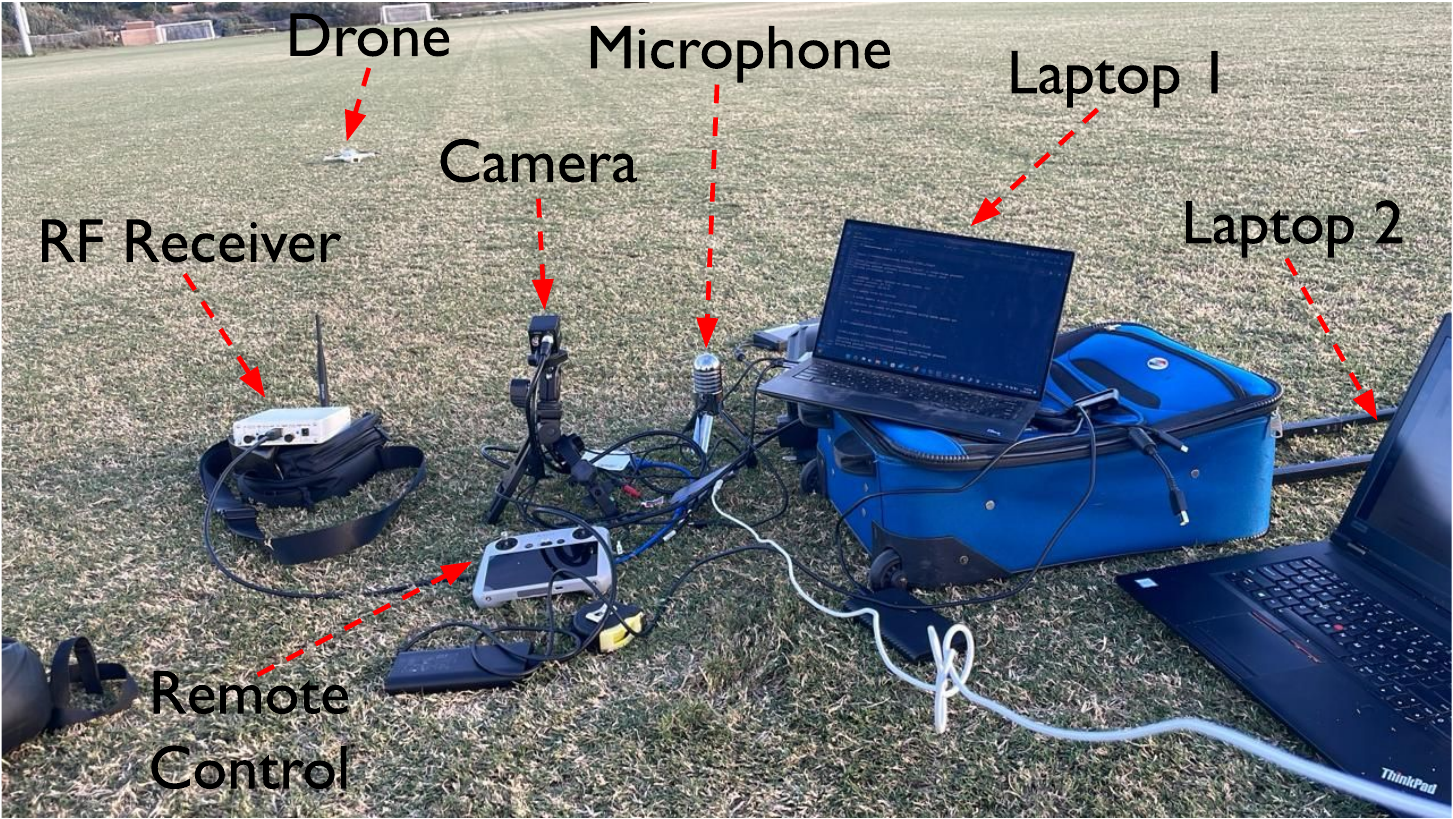}
    \captionsetup{font=small}
    \vspace{-0.4cm} 
    \caption{Experimental setup for synchronizing sensor data.}
    \label{fig:datacollection}
    \vspace{-0.35cm} 
\end{figure}

In our study, we employed a comprehensive sensing setup to analyze and detect drones under various conditions. Specifically, we focused on two drone models: DJI Mini 2 \cite{dji2023mini2se} and DJI Mini 3 Pro \cite{dji2023minipro}. These models were selected due to their distinct characteristics, including differences in sound profiles, size, transmission range, and flight speed, enabling a robust evaluation of the detection system in diverse operational scenarios. 
The data collection process relied on audio, video, and RF sensors into our experimental setup (Figure~\ref{fig:datacollection}). Each sensor type was selected to complement the others, ensuring precise and synchronized data acquisition across scenarios. 
The specifications of the devices are detailed in 
the Appendix~\ref{appendix A}.

\noindent \tikz[baseline=(char.base)]\node[shape=circle,draw,inner sep=1pt] (char) {1}; \noindent\textbf{Audio Sensor.}~ The Samson Meteor condenser microphone was used to capture drone acoustic signatures. It features a cardioid polar pattern with a sensitivity of -33 dB/Pa and records at 44.1 kHz sampling rate with 16-bit resolution \cite{samsontech2024meteormic}.

\noindent \tikz[baseline=(char.base)]\node[shape=circle,draw,inner sep=1pt] (char) {2}; \noindent\textbf{Video Sensor.}~The Marshall CV-505 camera was selected for its compact design and high resolution \cite{marshall2024cv505mb}. It captures 10-second video clips at 640 × 640 resolution with 30 frames per second (fps). 

\noindent \tikz[baseline=(char.base)]\node[shape=circle,draw,inner sep=1pt] (char) {3}; \noindent\textbf{RF Sensor.}~The USRP B210 software-defined radio (SDR), equipped with a VERT2450 
antenna, was configured to operate in the 2.4 GHz ISM band. The system recorded in-phase and quadrature (I/Q) data at 55 MS/s, with a center frequency of 2.4415 
GHz. 

\vspace{-0.2cm} 
\subsection{Experimental Conditions}
Data was collected under diverse environmental and operational conditions to assess the system's robustness and real-world applicability, ensuring reliability even when some sensors were partially degraded or less effective.

\noindent \textbf{Light Conditions.}~Data was collected during \textit{daylight} and \textit{sunset} to evaluate the system’s performance under different lighting conditions. These scenarios tested the video sensor’s effectiveness under varying visibility and assessed how the system adapts when the video sensor is less effective, relying more on audio and RF data for detection.


\noindent \textbf{Noise Levels.}~Experiments were conducted in environments with varying noise intensities. A \textit{stadium setting} provided moderate background noise from vehicles and people, while an \textit{urban area} introduced higher noise levels from aircraft and dense activity, challenging the system’s ability to isolate drone signals.


\noindent \textbf{Environmental Factors.}~ Data was collected in both line-of-sight (LOS) and non-line-of-sight (NLOS) conditions to evaluate the impact of visual and physical obstructions on sensor performance. In LOS conditions, the drone had a clear path to the sensors, ensuring optimal performance for video, audio, and RF detection. In NLOS conditions, obstacles like trees or walls partially or fully blocked the drone, affecting all sensors to varying degrees.


\vspace{-0.2cm} 
\subsection{Data Collection Procedure}

The data collection process was designed to capture high-quality sensor data across diverse scenarios. Locations were carefully selected based on environmental characteristics such as open spaces, interference sources, and background noise levels. Optimal sensor placements and drone flight paths were determined to ensure comprehensive coverage. A shared timestamp system synchronized all sensors to maintain seamless data alignment.
\begin{figure}[h]
    \centering
    \vspace{-0.35cm} 
    \includegraphics[scale=0.28]{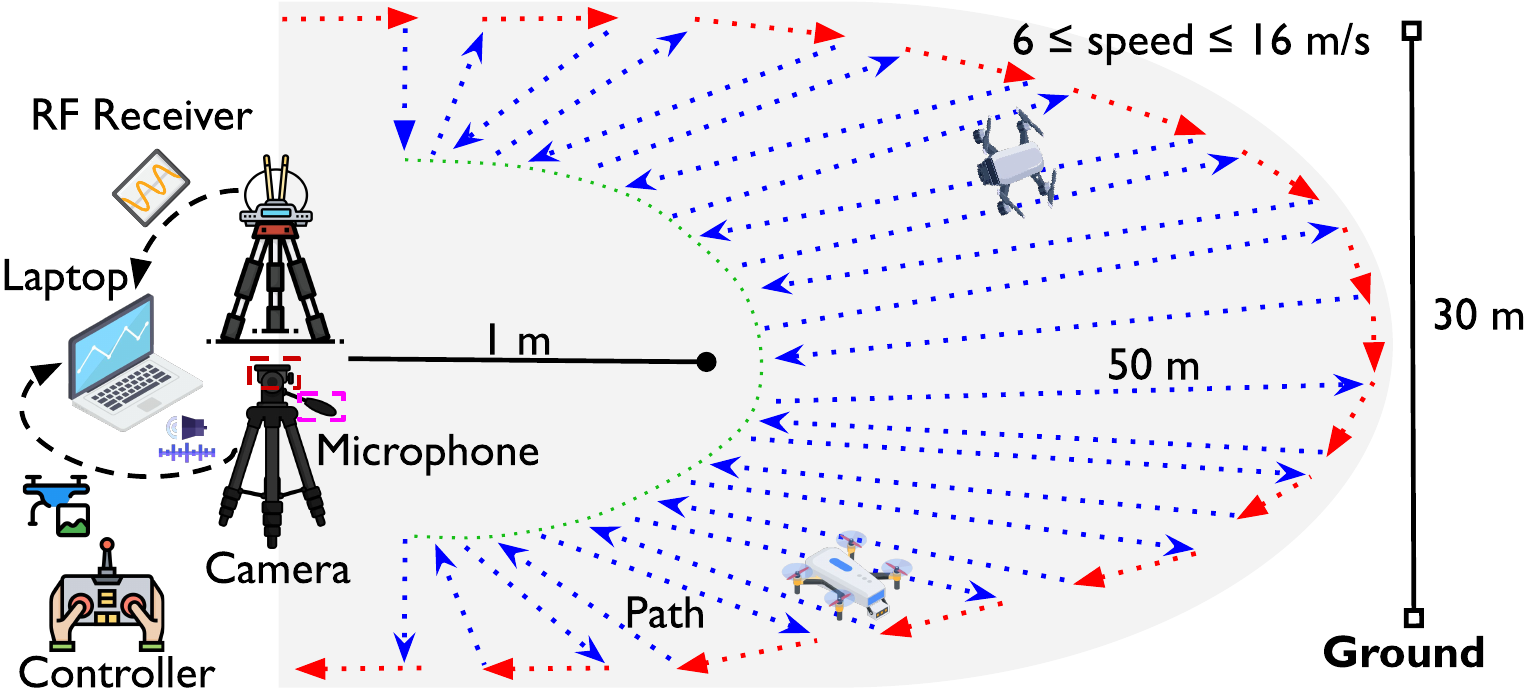}
    \vspace{-0.4cm}
    \captionsetup{font=small} 
    \caption{Data acquisition method from drones.}
    \label{fig:data_acquisition}
    \vspace{-0.4cm} 
\end{figure}

Drone flights included a variety of scenarios to replicate real-world conditions. Hovering scenarios maintained fixed altitudes between 1 and 30 meters (Figure~\ref{fig:data_acquisition}) to ensure stable data collection. Linear flights simulated common applications like surveillance and delivery by following straight paths at varying speeds. Horizontal movements covered distances from 1 to 50 meters, capturing how proximity affects sensor data. Agile maneuvers introduced complexity through sharp turns, circular patterns, and other dynamic movements.
These scenarios were repeated under both daylight and sunset conditions, as well as in urban and non-urban settings, incorporating both LOS and NLOS configurations—where the drone operated behind obstacles like walls or trees—to evaluate sensor performance across varying environmental and lighting conditions.

\vspace{-0.2cm} 
\section{Sensor Fusion for Drone Detection}
\label{sensor fusion}

\subsection{Data Pre-processing}

\tikz[baseline=(char.base)]\node[shape=circle, fill=black, inner sep=1pt, text=white] (char) {1}; \noindent \textbf{Data Segmentation.}~For our sensor fusion model, we segmented data into 0.25-second intervals to enhance real-time processing and ensure precise synchronization across audio, video, and RF modalities. 
Each audio segment corresponds to seven frames of video and one frame of RF spectrograms, creating a comprehensive dataset for analysis. This approach balances temporal resolution and computational efficiency while aligning with the characteristics of drone signatures from three sensors.

\begin{figure}[h]
    \centering
        \includegraphics[scale=0.152]{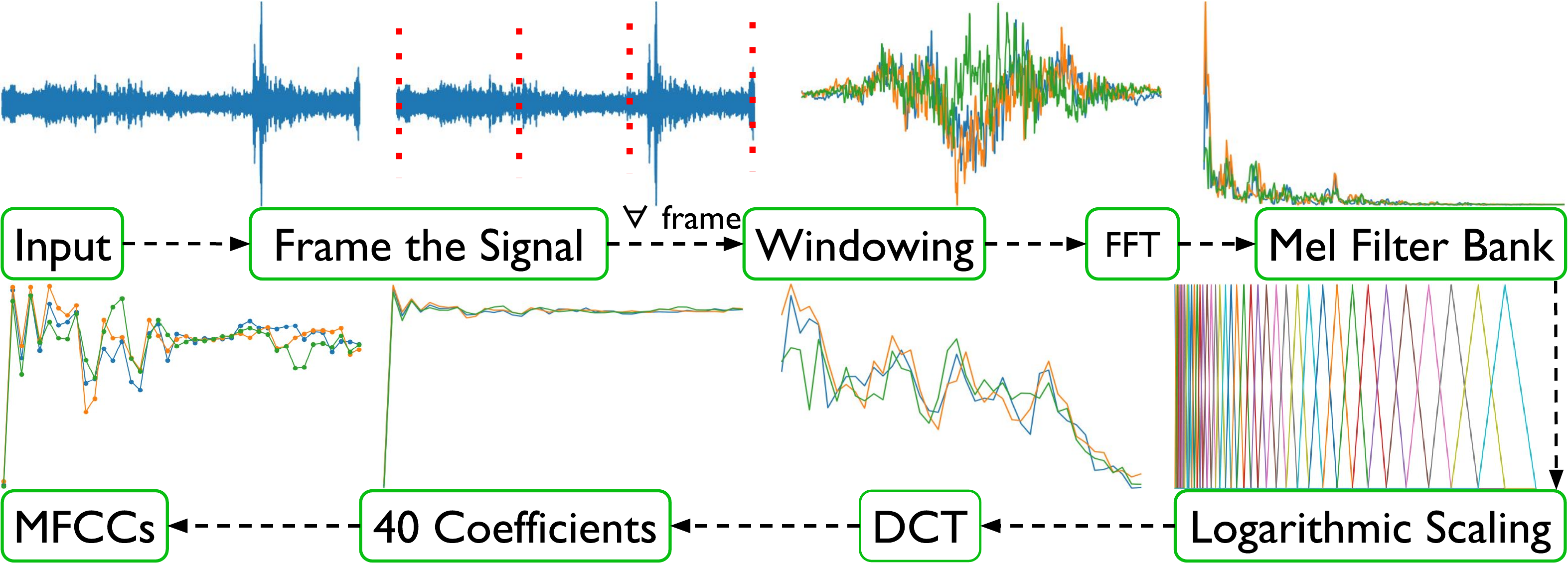}
    
    \captionsetup{font=small}
    \vspace{-0.4cm} 
    \caption{Audio feature extraction process.}
    \label{fig:audio_feature_extraction}
    \vspace{-0.47cm} 

\end{figure}

\noindent \tikz[baseline=(char.base)]\node[shape=circle, fill=black, inner sep=1pt, text=white] (char) {2}; \textbf{Audio Pre-processing.}~
We use Mel Frequency Cepstral Coefficients (MFCCs) to extract features crucial for drone detection, leveraging their robustness in capturing the spectral characteristics of audio signals.
MFCCs are particularly effective in modeling the unique acoustic signatures of drones, as supported by prior research \cite{anwar2019machine, ohlenbusch2021robust, alla2024sound}.  Each audio segment is divided into 40 frames, with 40 MFCCs computed per frame.  The extraction process, as shown in Figure \ref{fig:audio_feature_extraction}, begins with windowing to minimize boundary discontinuities and ensure smooth transitions between frames. Fast Fourier Transform (FFT) is then applied to analyze the frequency components of the signal. A Mel filter bank focuses on critical auditory frequencies, and logarithmic scaling models loudness perception. Finally, a Discrete Cosine Transform (DCT) decorrelates the energy distribution across the Mel filters, resulting in 40 MFCCs that effectively capture the acoustic signature of drones.

\noindent \tikz[baseline=(char.base)]\node[shape=circle, fill=black, inner sep=1pt, text=white] (char) {3}; \textbf{Video Pre-processing.}~We began by decomposing the recorded video clips into individual frames, converting the continuous video stream into a discrete sequence of images for frame-by-frame analysis. Each extracted frame was resized to a uniform resolution of 112×112 pixels, ensuring consistent input dimensions and reducing computational complexity for subsequent deep learning models.   Furthermore, to capture short-term temporal dynamics within the video data while maintaining computational efficiency, we employed a frame stacking approach.  Specifically, consecutive sequences of seven frames were stacked along a new dimension to form a single input sample for the video-based unimodal models. 

\noindent \tikz[baseline=(char.base)]\node[shape=circle, fill=black, inner sep=1pt, text=white] (char) {2}; \textbf{RF Pre-processing.}~For RF data, raw I/Q samples are converted into spectrograms, ensuring a high-resolution representation of the temporal and spectral characteristics essential for drone detection. The transformation employs the Short-Time Fourier Transform (STFT), defined as:
\vspace{-0.1cm}
\begin{equation}
S(t, f) = \sum_{n=0}^{N-1} x[n] \cdot w[n-t] \cdot e^{-j2\pi fn/N},
\end{equation}
\vspace{-0.2cm}

\noindent where \(x[n]\) is the I/Q data, \(w[n]\) is the Hamming window function, \(t\) is the time index, and \(f\) is the frequency.
Each spectrogram is treated as an image and automatically resized to 112$\times$112 pixels.

\vspace{-0.2cm} 
\subsection{Data Augmentation}
\label{Data Augumentation}

\begin{figure}[htbp]
    \centering
    \vspace{-0.2cm} 
    \includegraphics[width=0.3\textwidth]{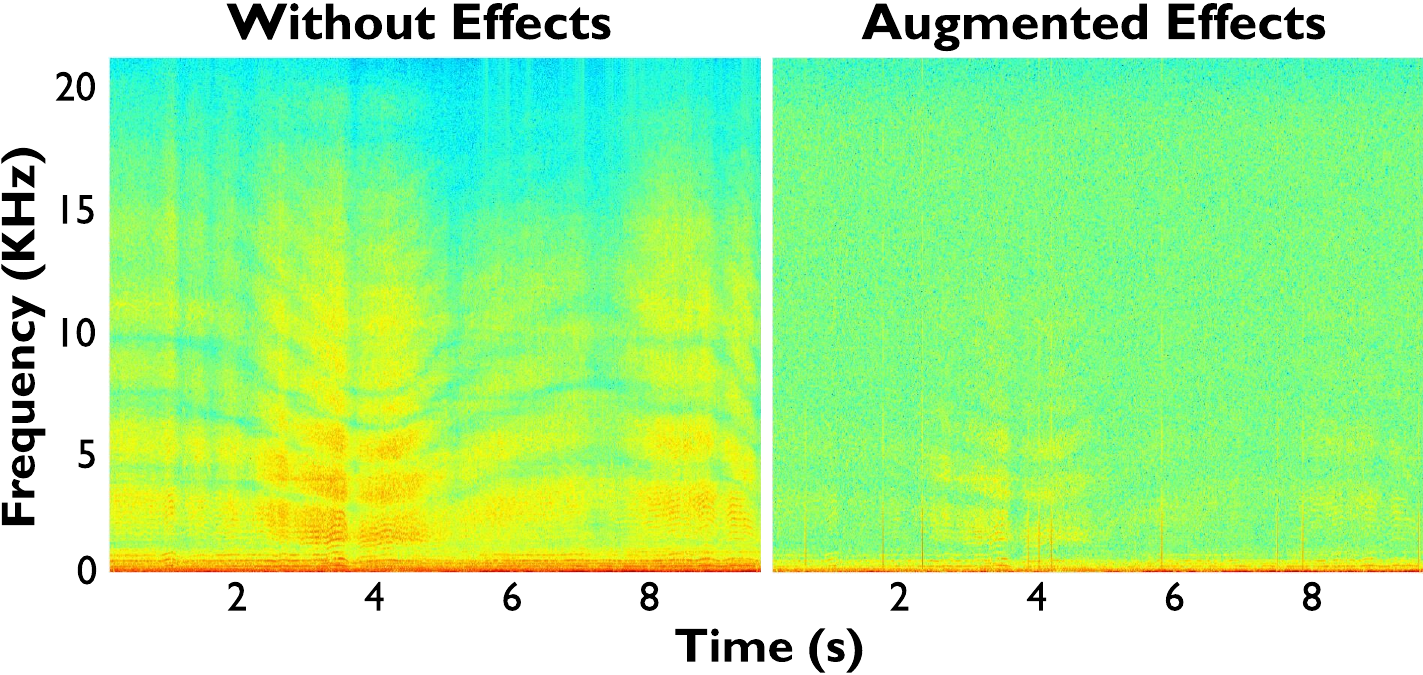}
    \vspace{-0.4cm} 
    \caption{T-F representation of drone sound without/with background noise and volume scaling effects.}
    \vspace{-0.35cm} 
    \label{fig: audio augumentation_effects}
\end{figure}

\noindent \tikz[baseline=(char.base)]\node[shape=circle,draw,inner sep=1pt] (char) {1}; \textbf{Audio Augmentation.}~
We applied a range of audio augmentation techniques  to evaluate the system’s robustness against noise, signal variability, and environmental factors. These techniques included adding \textit{background noise} to simulate conditions where drone sounds might be obscured by other noises, such as traffic, wind, or human activity.  \textit{Harmonic distortions} were introduced to emulate non-linear variations in microphone response, while \textit{pitch shifting} modified the frequency to account for changes in drone motor speed or Doppler effects. \textit{Clicks} were generated to represent sudden disturbances, such as electrical noise or short sound bursts caused by interference. \textit{Mono conversion} was applied to simulate audio input from a single-channel sensor type that is different from multi-channel one. Finally, \textit{volume scaling} adjusted audio levels to reflect varying recording distances and sound intensities. 
The effects of these augmentations are shown in Figure \ref{fig: audio augumentation_effects}. These transformations enriched the dataset by simulating real-world acoustic variations, testing the system’s resilience 
under challenging conditions.

The impact of these audio augmentations on the original signal was evaluated using the Mel Cepstral Distortion (MCD) metric \cite{saldanha2022data}. MCD quantifies the spectral dissimilarity between the original and augmented audio, providing a quantitative measure of the introduced noise and distortion. It is calculated using the following formula in decibels (dB):
\vspace{-0.25cm}


\begin{equation}\small
    MCD
= \frac{10\sqrt{2}}{\ln(10)} \cdot \frac{1}{T}
\sum_{t=1}^{T}
\sqrt{
  \sum_{m=1}^{n_{\text{coeffs}}}
  \bigl(MFCC_{real}(m,t) - MFCC_{aug}(m,t)\bigr)^{2}
},
\end{equation}
where \(MFCC_{real}(m, t)\) and \(MFCC_{aug}(m, t)\) represent the \(m\)-th MFCC coefficient at time frame \(t\) for the real and augmented audio signals, respectively. \(T\) is the total number of time frames, and \(n_{\text{coeffs}}\) is the number of MFCC coefficients.
Lower MCD values indicate minimal distortion, while higher values reflect greater spectral dissimilarity between the original and augmented signals.

\begin{figure}[htbp]
    \centering
    \vspace{-0.2cm} 
    \includegraphics[width=0.47\textwidth]{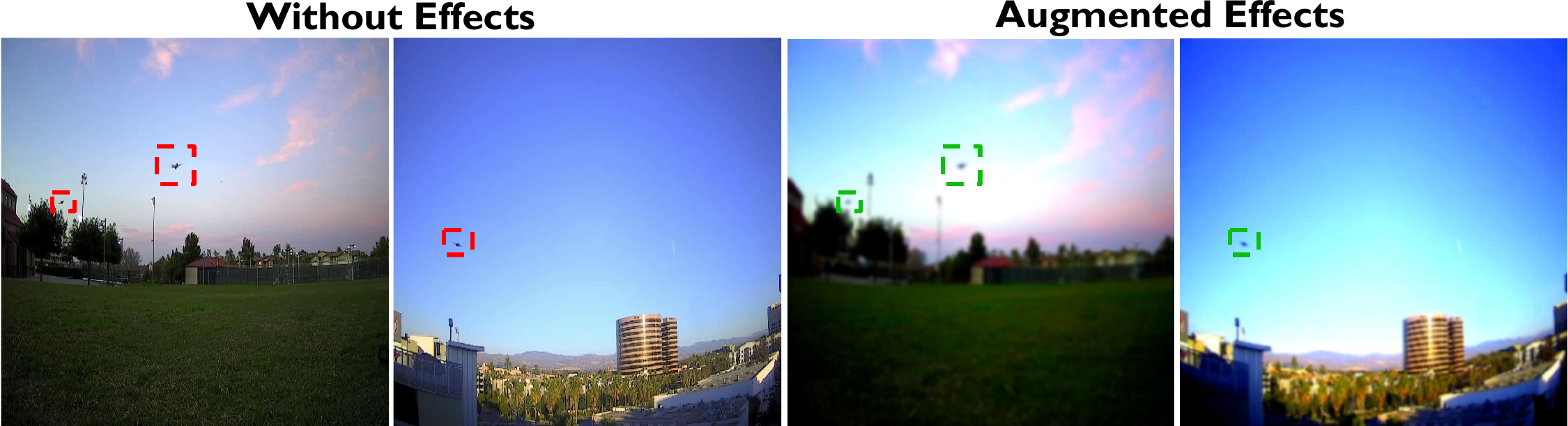}
    \vspace{-0.4cm} 
    \caption{Drone image without/with Gaussian blur and color jitter.}
    \vspace{-0.35cm}
    \label{fig: image_augumentation}
\end{figure}

\noindent \tikz[baseline=(char.base)]\node[shape=circle,draw,inner sep=1pt] (char) {2}; \textbf{Visual Augmentation.}~Several augmentation techniques were applied to the video data to enhance the dataset’s diversity and better reflect real-world conditions. These techniques included adding \textit{random noise} to replicate sensor artifacts by introducing pixel-level distortions, representing degraded image quality.  \textit{Horizontal flipping} changed the frame orientation to account for variations in perspective, while \textit{rotation} introduced angle shifts to simulate different viewpoints. \textit{Color jitter} adjusted brightness and contrast to recreate diverse lighting conditions, such as overexposure or shadows. \textit{Gaussian blur} softened the images to mimic motion blur or out-of-focus visuals. Finally, \textit{salt-and-pepper noise} added random white and black pixels to simulate transmission errors or sensor defects, degrading visual clarity. As shown in Figure \ref{fig: image_augumentation}, these transformations effectively diversified the dataset by incorporating a range of challenging conditions.

The impact of these visual augmentations on the original data was evaluated using the Structural Similarity Index (SSIM) \cite{akbiyik2023data}. SSIM provides a perceptual similarity metric between the original and augmented video frames, with values ranging from 0 to 1, where 1 indicates perfect similarity and lower values signify increasing distortion. The SSIM calculation is given by Eq \ref{equation 2}:
\begin{equation}\small
\label{equation 2}
SSIM(x, y) = \frac{(2\mu_x\mu_y + C_1)(2\sigma_{xy} + C_2)}{(\mu_x^2 + \mu_y^2 + C_1)(\sigma_x^2 + \sigma_y^2 + C_2)},
\end{equation}
where \( \mu_x, \mu_y \) represent the luminance of the original image (\(x\)) and and the augmented image (\(y\)),  \( \sigma_x \) and \(\sigma_y\) denote their contrast, and \( \sigma_{xy} \) represents their structural similarity. The constants \(C_1\) and \(C_2\) stabilize the division to avoid numerical instabilities.


\noindent \tikz[baseline=(char.base)]\node[shape=circle,draw,inner sep=1pt] (char) {3}; \textbf{RF Augmentation.}~
Two distinct data collection scenarios were designed: \textit{in-the-wild (ITW)} and \textit{over-the-air (OTA)}.
\textit{In ITW scenario}, RF signals were recorded in a controlled laboratory using a USRP B210, capturing real-world interference from Wi-Fi devices and IoT sensors operating in the 2.4 GHz band. This dataset realistically represents RF environments where multiple devices share the same frequency spectrum. \textit{In the OTA scenario}, drone signal transmissions were recorded in an outdoor environment, capturing real-world RF signals with natural interference. For increased interference, a USRP B200 Mini was configured as a jammer, injecting artificial noise into the 2.4 GHz band. Gaussian and uniform noise were digitally generated and transmitted at different bandwidths (e.g., 10 MHz and 20 MHz), simulating multiple overlapping transmissions. 
\begin{figure}[htbp]
    \centering
    \vspace{-0.25cm}
    \includegraphics[width=0.33\textwidth]{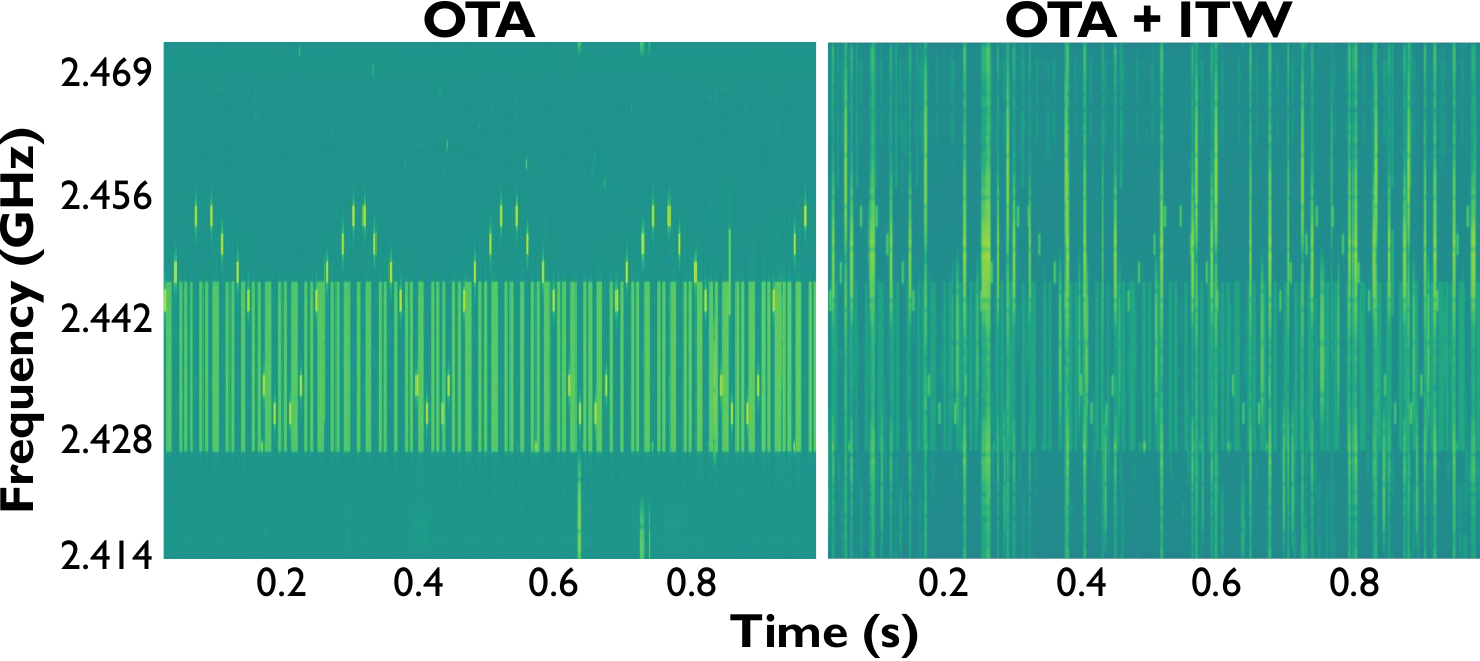}
    \vspace{-0.4cm}
    \caption{RF Spectrograms with OTA and OTA+ITW cases.}
    \vspace{-0.4cm}
    \label{fig:rf_augumentation}
\end{figure}

We enhanced the dataset’s complexity and diversity by combining ITW and OTA data through additive operations. This process involved mixing drone communication signals collected in the OTA scenario with laboratory-collected noise from the ITW scenario. As shown in Figure \ref{fig:rf_augumentation}, OTA scenarios display distinct drone communication patterns, while OTA+ITW scenarios show these patterns increasingly obscured by background noise. This combination reflects real-world challenges and enables a comprehensive evaluation of detection robustness.

The level of noise introduced during RF data augmentation was quantified using the Signal-to-Noise Ratio (SNR) in dB \cite{shen2023toward}. SNR measures the signal power relative to the noise power, effectively indicating the level of interference added.  It is calculated as:
\vspace{-0.3cm}

\begin{equation}\small
SNR = 10 \cdot \log_{10} \left( \frac{P_{signal}}{P_{noise}} \right),
\end{equation}
where \(P_{signal}\) represents the power of the drone communication signal and \(P_{noise}\) denotes the power of the injected noise. Lower SNR values indicate higher levels of noise relative to the signal. 

\vspace{-0.2cm}
\subsection{Unimodal Feature Extraction}
\begin{figure*}[h]
    \centering
    \includegraphics[width=0.7\textwidth]{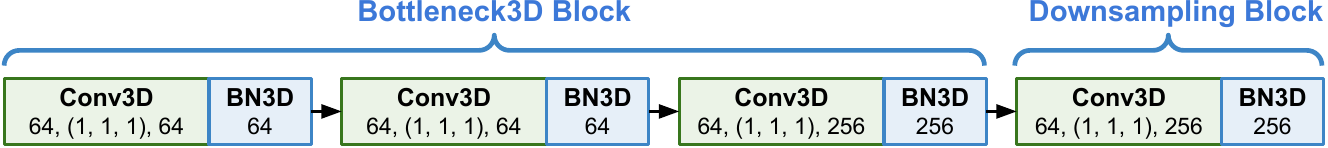}
    \vspace{-0.4cm} 
    \captionsetup{font=small}
    \caption{The architecture breakdown of the bottleneck3D and downsample layers for '\text{Block\_1}' in Table \ref{tab:resnet_model}.}
    \label{fig:resnet_blocks}
    \vspace{-0.4cm} 

\end{figure*}

\subsubsection{Models for Audio Feature Extraction}
The pre-processing step of raw audio data generates 2-dimensional vectors with a shape of \((1,1600)\) from various audio recordings. These 2D vectors are then utilized to extract features relevant to UAV detection. This feature extraction process can be systematically achieved using a ML model. CNNs have proven to be highly effective in learning features from audio, visual, and RF data. For our specific use case, we employ two different state-of-the-art CNN architectures, with further customization to accommodate the unique nature and shape of our audio data. In the following, we detail the architectural specifications and adjustments that we have made for each model:

\noindent\textbf{(i) LeNet-based Architecture.}~ 
As detailed in Table \ref{tab:lenet_model}, this architecture is composed of four feature extraction blocks -- each block includes a convolutional (Conv) layer, a batch normalization (BN) layer, and a max pooling (MaxPool) layer. The number of feature maps in these blocks ranges from 8 to 32, which helps to balance the model's complexity and computational efficiency, making it both compact and lightweight. To accommodate the Conv2D layers in our CNN, we perform a dimension reshaping of the audio vector. Initially, the audio vector has dimensions of \((1,1600)\). We reshape this vector to a size of \((1,40,40)\). This transformation is crucial as it enables the use of 2D convolutional layers, which are better suited for extracting spatial hierarchies and intricate patterns from the data compared to 1D convolutional layers.

\begin{table}[ht]
\centering
\captionsetup{font=small}
\caption{Architecture specifications of audio-LeNet.}
\vspace{-0.4cm}
\label{tab:lenet_model}
\resizebox{\columnwidth}{!}{ 
\begin{tabular}{llll} 
\hline
\textbf{Block} & \textbf{Neural Layers Operations} & \textbf{Feature map} & \textbf{Activation} \\ 
\hline
Block\_1 & Conv(3x5x5), BN(8), MaxPool(2) & 8 & ReLU \\
Block\_2 & Conv(8x3x3), BN(16), MaxPool(2) & 16 & ReLU \\
Block\_3 & Conv(16x3x3), BN(24), MaxPool(2) & 24 & ReLU \\
Block\_4 & Conv(24x3x3), BN(32), MaxPool(2) & 32 & ReLU \\ 
\hline
Classifier & Fully-Connected(32, 2) & 32 & Sigmoid() \\
\hline
\end{tabular}
}
\vspace{-0.45cm}
\end{table}

\noindent\textbf{(ii) VGG-based Architecture.}~We employ a more complex and over-parameterized backbone for audio feature extraction, utilizing the VGGNet architecture \cite{simonyan2014very}, which was originally designed for visual data processing. Specifically, we adopt a VGG-19 variant to process audio data in the same manner as previously explained for the LeNet model. The backbone of VGG-19 depicts a wide and deep architecture composed of 16 convolutional layers with 3x3 kernels, sometimes followed by max pooling layers. The number of feature maps in these layers ranges from 64 to 512, allowing the model to capture a broad spectrum of audio features with high precision.
We modify the classifier part of the VGG-19 architecture to optimize the model for audio data.
Specifically, we remove the two fully-connected layers originally included in the VGG-19 model and replace them with a single, less parameterized fully-connected layer. This adjustment not only reduces the model's complexity but also enhances processing time and facilitates seamless training. Our final audio-VGG model architecture is detailed in Table \ref{tab:vgg_model}. To simplify the reading, we group the layers by blocks, each containing two convolutional layers with specifications on the number and spatial dimensions of their kernels.
\vspace{-0.35cm}
\begin{table}[ht]
\centering
\captionsetup{font=small}
\caption{Architecture specifications of audio-VGG.}
\vspace{-0.4cm}
\label{tab:vgg_model}
\resizebox{\columnwidth}{!}{ 
\begin{tabular}{llll} 
\hline
\textbf{Block} & \textbf{Neural Layers Operations} & \textbf{Feature map} & \textbf{Activation} \\ 
\hline
Block\_1 & Conv(1x3x3), Conv(64x3x3), MaxPool(2) & 64, 64 & ReLU \\
Block\_2 & Conv(64x3x3), Conv(128x3x3), MaxPool(2) & 128, 128 & ReLU \\
Block\_3 & Conv(128x3x3), Conv(256x3x3) & 128, 256 & ReLU \\
Block\_4 & Conv(256x3x3), Conv(256x3x3), MaxPool(2) & 256, 256 & ReLU \\
Block\_5 & Conv(256x3x3), Conv(512x3x3) & 256, 512 & ReLU \\
Block\_6 & Conv(512x3x3), Conv(512x3x3) & 512, 512 & ReLU \\
Block\_8 & Conv(512x3x3), Conv(512x3x3), MaxPool(2) & 512, 512 & ReLU \\ 
\hline
Classifier & Fully-Connected(512, 2) & 512 & Sigmoid() \\
\hline
\end{tabular}
}
\vspace{-0.55cm}
\end{table}

\subsubsection{Models for Visual and RF Feature Extraction}

Visual data are pre-processed and segmented into sequences of frames featuring the presence/absence of the drone. In parallel, RF data (i.e., spectrograms derived from I/Q data transformation) are also organized into frame-like inputs—typically one spectrogram 'frame' per segment. Since we're more interested in detecting the presence/absence of the drone, we consider a binary classification task where labels are assigned to both video and RF segments based on the underlying frames. Formally, let \(Y\) be the binary label assigned to a video or RF sequence, and \(y_i\) be the binary label of the \(i\)-th frame:
\vspace{-0.2cm}

\begin{equation}
Y = \max(y_1, y_2, \ldots, y_n),
\end{equation}
where \(i = 1, 2, \ldots, n\), and \(n\) is the total number of frames. The \(\max\) function returns 1 if \emph{at least one} frame is labeled 1, and 0 otherwise.

The pre-processing of video and RF frames results in 4D vectors with a shape of \(X_{\text{shape}} = (N, C, H, W)\), where \(N\) is the number of frames, \(C\) is the number of color channels per frame, and \(H\) and \(W\) are the height and width, respectively. By incorporating the \(N\) dimension for stacking a set of frames, both temporal and spatial dependencies can be learned from the sequences. Unlike single-frame classification, which uses 3D vectors, video classification requires specific ML models with filters extended into an additional dimension to train on both spatial and temporal data. Therefore, we utilize 3D CNN models featuring convolutional filters in three-dimensional space. We have selected two of the most efficient and effective state-of-the art CNN models for visual tasks: MobileNet-v2 and ResNet-10 \cite{howard2017mobilenets, hara2018can}. Since these models were originally proposed for single-frame classification, we applied a further transformation to map all their layers in the three-dimensional space. Further details on the architectural specifications of each model are provided below:

\noindent\textbf{(i) ResNet-based Architecture.}~ResNet models are highly successful architectures for learning visual features \cite{he2016deep}. We adopt a lightweight ResNet-10 variant with fewer weights and a simpler design, incorporating both bottleneck and downsampling blocks. As detailed in Table \ref{tab:resnet_model}, the backbone of ResNet-10 consists of 10 convolutional layers, organized into bottleneck3D and downsample blocks with 3x3x3 and 1x1x1 kernels, followed by batch normalization layers, as illustrated in Figure \ref{fig:resnet_blocks}. The number of feature maps in these layers ranges from 64 to 2048, enabling the model to capture a wide range of video features with high precision. In the classifier block, we apply spatio-temporal average pooling with a kernel size of $(N, 4, 4)$, where the number of frames is $N=7$. This pooling layer performs downsampling along the spatial dimensions (depth, height, and width) by averaging the values over the pooling window (i.e., $(7, 4, 4)$) for each channel of the input. 
For RF data, \(N\!=\!1\), so the same pooling simply reduces the spatial dimensions.
Finally, a fully connected layer processes the output of the pooling operation and produces a single classification score for the $N$ frames, which is then translated into a predicted label for the entire video sequence.

\begin{table}[ht]
\centering
\vspace{-0.35cm}
\captionsetup{font=small}
\caption{Architecture specifications of ResNet-10.}
\vspace{-0.4cm}
\label{tab:resnet_model}
\resizebox{\columnwidth}{!}{ 
\begin{tabular}{llll} 
\hline
\textbf{Block} & \textbf{Neural Layers Operations} & \textbf{Fmap} & \textbf{Activation} \\ 
\hline
Conv\_1 & Conv2D(3x7x7),~MaxPool(3), BN(64) & 64 & ReLU \\
Block\_1 & Bottleneck3D(64, 256), Downsample(64, 256) & 256 & ReLU \\
Block\_2 & Bottleneck3D(256, 512), Downsample(256, 512) & 512 & ReLU \\
Block\_3 & Bottleneck3D(512, 1024), Downsample(512, 1024) & 1024 & ReLU \\
Block\_4 & Bottleneck3D(1024, 2048), Downsample(1024, 2048) & 2048 & ReLU \\ 
\hline
Classifier & AvgPooling3D(N, 4, 4), Fully-Connected(2048, 2) & 2 & Sigmoid() \\
\hline
\end{tabular}
}
\vspace{-0.3cm}
\end{table}

\noindent\textbf{(ii) MobileNet-based Architecture.}~ 
MobileNet models were initially introduced to enhance processing efficiency, particularly on resource-constrained hardware devices, while maintaining high performance on visual tasks \cite{howard2017mobilenets}. As such, MobileNet variants are lightweight and compact, with fewer trainable weights and simpler operations, making them ideal for real-time applications. Consequently, to meet the real-time requirements of our UAV detection systems, we utilize the MobileNet architecture to learn visual features from video and RF sequences. 
We have further modified the original MobileNet by transforming the MBConv2D blocks into MBConv3D blocks to handle the processing of 4D vectors that represent sequences. Table \ref{tab:mobilenet_model} outlines the architectural specifications of the MobileNet3D model. It features a sequence of MBConv3D layers, which are composed of a typical Conv3D layer followed by batch normalization and Rectified Linear Unit (ReLU) activation. The classifier block involves a spatio-temporal average pooling with a kernel size of $(28, 1, 1)$ and a fully connected layer with sigmoid activation. For the single-frame RF input, the temporal dimension \(N\) is 1.

\begin{table}[ht]
\centering
\vspace{-0.25cm}
\captionsetup{font=small}
\caption{Architecture specifications of MobileNet.}
\vspace{-0.4cm}
\label{tab:mobilenet_model}
\resizebox{\columnwidth}{!}{ 
\begin{tabular}{llll} 
\hline
\textbf{Block} & \textbf{Neural Layers Operations} & \textbf{Fmap} & \textbf{Activation} \\ 
\hline
Conv\_1 & Conv2D(3x3x3x3), BN(32) & 32 & ReLU \\
Block\_1 & MBConv3D(32, 64),~MBConv3D(64, 128),~ & 128 & ReLU \\
Block\_2 & MBConv3D(128, 128),~MBConv3D(128, 256) & 256 & ReLU \\
Block\_3 & MBConv3D(256, 256),~MBConv3D(256, 512) & 512 & ReLU \\
Block\_4 & MBConv3D(512, 512),~MBConv3D(512, 512), & 512 & ReLU \\
Block\_5 & MBConv3D(512, 512),~MBConv3D(512, 512) & 512 & ReLU \\
Block\_6 & MBConv3D(512, 512),~MBConv3D(512, 1024) & 1024 & ReLU \\
Block\_7 & MBConv3D(1024, 1024) & 1024 & ReLU \\ 
\hline
Classifier & AvgPooling3D(N, 1, 1), Fully-Connected(256, 2) & 2 & Sigmoid() \\
\hline
\end{tabular}
}
\vspace{-0.5cm}
\end{table}

\vspace{-0.2cm}
\subsection{Fusion Strategies}

\begin{figure}[h]
    \centering
    \includegraphics[width=0.478\textwidth]{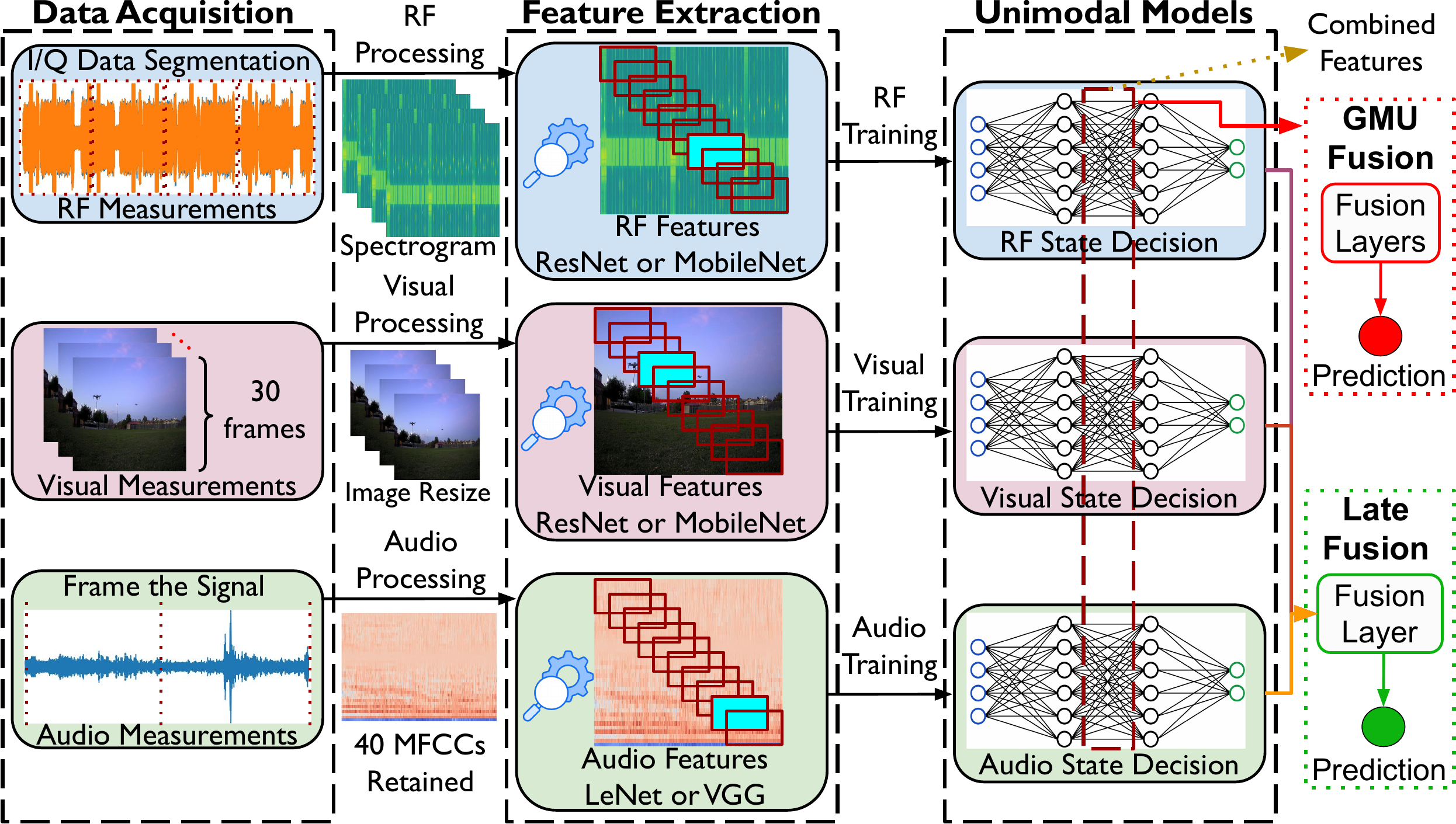}
    \vspace{-0.7cm}
    \caption{\IW Framework.}
    \label{fig:detection_framework}
    \vspace{-0.6cm}
\end{figure}

The fusion techniques are key components of the \IW framework, as shown in Figure~\ref{fig:detection_framework}. In this pipeline, audio, visual, and RF data are collected, pre-processed, and passed through modality-specific models (LeNet or VGG-19 for audio, ResNet-10 or MobileNet for video and RF). The resulting unimodal features or decisions are then combined via GMU or Late Fusion to yield the final drone prediction. The following subsections describe the fusion strategies.
\vspace{-0.5cm}

\subsubsection{Late Fusion}
Late Fusion is one straightforward way to fuse the output predictions calculated by each unimodal model. 
The computations involving each unimodal model should be performed independently, and only their final output predictions (i.e., the output of the classifier block) are considered for the fusion. We note that the output predictions are also probability distributions produced by unimodal models in the case of classification. Thus, Late Fusion is the most interpretable method, operating on the highest level of unimodal output data. Algorithm \ref{alg:late_fusion} reports details of the steps involved in Late Fusion. Let $\mathbf{y}_{Aud}$, $\mathbf{y}_{Vis}$, $\mathbf{y}_{RF}$ be the output predictions (i.e., probabilities) of absence/presence of UAV computed by audio, video and RF unimodal models, respectively. Given these outputs, Late Fusion operates as follows:

\noindent \textbf{Step \noindent \tikz[baseline=(char.base)]\node[shape=circle,draw,inner sep=1pt] (char) {\textbf{1}};}: Output predictions are combined using normalized fusion weights, $\alpha$, $\beta$, and $\gamma$, which determine the contribution of each output prediction to the final prediction. Fusion weights are learned during training based on the importance of each modality and are normalized as follows:
\vspace{-0.14cm}
\begin{equation}
    \alpha' = \frac{\alpha}{\alpha + \beta + \gamma}, \beta' = \frac{\beta}{\alpha + \beta + \gamma}, \gamma' = \frac{\gamma}{\alpha + \beta + \gamma}.
    \vspace{-0.15cm}
\end{equation}

\noindent \textbf{Step \noindent \tikz[baseline=(char.base)]\node[shape=circle,draw,inner sep=1pt] (char) {\textbf{2}};}: The final output prediction, $\mathbf{y}$, is obtained by applying a weighted sum on the the unimodal output predictions $\mathbf{y}_{Aud}$, $\mathbf{y}_{Vis}$, and $\mathbf{y}_{RF}$ using normalized weights as follows:
\vspace{-0.15cm}
\begin{equation}
    \mathbf{y} = \alpha' \mathbf{y}_{Aud} + \beta' \mathbf{y}_{Vis} + \gamma' \mathbf{y}_{RF}.
    \vspace{-0.1cm}
\end{equation}

\vspace{-0.25cm}
\begin{algorithm}
\small
\captionsetup{font=small}
\caption{Late Fusion for audio-video data.}
\label{alg:late_fusion}
\begin{algorithmic}[1]
\REQUIRE Input predictions $\mathbf{y}_{Aud}$, $\mathbf{y}_{Vis}$, $\mathbf{y}_{RF}$
\REQUIRE Fusion weights $\alpha, \beta, \gamma$
\STATE Normalize the fusion weights: $\alpha' = \frac{\alpha}{\alpha + \beta + \gamma}$, $\beta' = \frac{\beta}{\alpha + \beta + \gamma}$, $\gamma' = \frac{\gamma}{\alpha + \beta + \gamma}$
\STATE Compute the fused prediction: $\mathbf{y} = \alpha' \mathbf{y}_{Aud} + \beta' \mathbf{y}_{Vis} + \gamma' \mathbf{y}_{RF}$ 
\RETURN $\mathbf{y}$
\end{algorithmic}
\normalsize 
\end{algorithm}
\vspace{-0.3cm}

\vspace{-0.2cm}
\subsubsection{Gated Multimodal Unit (GMU) Fusion}
GMU Fusion is designed to effectively fuse information from multiple modalities by learning each contribution to the final prediction. 
Let
\( \mathbf{x}_{\mathit{Aud}}\), \( \mathbf{x}_{\mathit{Vis}}\), and \(\mathbf{x}_{\mathit{RF}}\), be the inputs of the GMU Fusion. Typically, these inputs are selected from the output intermediate feature maps of the unimodal models. Since the last output feature map (the input feature map of the classifier layer) usually encapsulates high-level features that are semantically meaningful, we use them as inputs to the GMU Fusion operation. Hence, contribution scores will be learned and assigned to each modality based on high-abstracted and meaningful output feature maps. Algorithm \ref{alg:GMU} details the working mechanism of the GMU Fusion. 

\vspace{-0.2cm}
\begin{algorithm}
\small
\captionsetup{font=small}
\caption{GMU Fusion for audio-video data.}
\label{alg:GMU}
\begin{algorithmic}[1]
\REQUIRE Input vectors $\mathbf{x}_{Aud}$, $\mathbf{x}_{Vis}$, $\mathbf{x}_{RF}$
\STATE Compute the gating vector: $\mathbf{g} = \sigma(\mathbf{W}_g [\mathbf{x}_{Aud}, \mathbf{x}_{Vis}, \mathbf{x}_{RF}] + \mathbf{b}_g)$
\STATE Compute the transformed vectors:
\STATE \quad $\mathbf{y}_1 = \tanh(\mathbf{W}_y^1 \mathbf{x}_{Aud} + \mathbf{b}_y^1)$
\STATE \quad $\mathbf{y}_2 = \tanh(\mathbf{W}_y^2 \mathbf{x}_{Vis} + \mathbf{b}_y^2)$
\STATE \quad $\mathbf{y}_3 = \tanh(\mathbf{W}_y^3 \mathbf{x}_{RF} + \mathbf{b}_y^3)$
\STATE Compute the fused representation: $\mathbf{y} = \mathbf{g_{1}} \odot \mathbf{y}_1 + \mathbf{g_{2}} \odot \mathbf{y}_2 + \mathbf{g_{3}} \odot \mathbf{y}_3$
\RETURN $\mathbf{y}$
\end{algorithmic}
\normalsize 
\end{algorithm}
\vspace{-0.35cm}

\noindent \textbf{Step \noindent \tikz[baseline=(char.base)]\node[shape=circle,draw,inner sep=1pt] (char) {\textbf{1}};}: A \textit{gating vector} $\mathbf{g}$ is computed from the inputs \( \mathbf{x}_{\mathit{Aud}}\), \( \mathbf{x}_{\mathit{Vis}}\), and \(\mathbf{x}_{\mathit{RF}}\) using a sigmoid activation function. The gating vector $\mathbf{g}$ serves as a control knob for the contribution of each modality to the joint information from the fused data. Formally, the computation of the gated vector is defined as follows:
\vspace{-0.15cm}
\begin{equation}
    \mathbf{g} = \sigma(\mathbf{W}_g [\mathbf{x}_{Aud}, \mathbf{x}_{Vis}, \mathbf{x}_{RF}] + \mathbf{b}_g),
    \vspace{-0.15cm}
\end{equation}
where $\sigma$ is the sigmoid function, $\mathbf{W}_g$ are the weights, and $\mathbf{b}_g$ is the bias term. The concatenation of the input vectors $\mathbf{x}_{Aud}$, $\mathbf{x}_{Vis}$, and $\mathbf{x}_{RF}$ ensures that the gating mechanism considers learned features from three modalities (i.e., audio, video, and RF features).

\noindent \textbf{Step \noindent \tikz[baseline=(char.base)]\node[shape=circle,draw,inner sep=1pt] (char) {\textbf{2}};}: Each of the inputs $\mathbf{x}_{Aud}$,  $\mathbf{x}_{Vis}$, and $\mathbf{x}_{RF}$ independently undergoes a linear transformation followed by a $\mathbf{tanh}$ activation function. This step produces the transformed vectors $\mathbf{y}_1$, $\mathbf{y}_2$, $\mathbf{y}_3$:
\vspace{-0.1cm}
\begin{align}
    \mathbf{y}_1 = \tanh(\mathbf{W}_y^1 \mathbf{x}_{Aud} + \mathbf{b}_y^1),\\
    \mathbf{y}_2 = \tanh(\mathbf{W}_y^2 \mathbf{x}_{Vis} + \mathbf{b}_y^2),\\
    \mathbf{y}_3 = \tanh(\mathbf{W}_y^3 \mathbf{x}_{RF} + \mathbf{b}_y^3),
    \vspace{-0.15cm}
\end{align}
\vspace{-0.4cm}

\noindent where, $\mathbf{W}_y^1$, $\mathbf{W}_y^2$, and $\mathbf{W}_y^3$ are the weights, and $\mathbf{b}_y^1$, $\mathbf{b}_y^2$, and $\mathbf{b}_y^3$ are the bias terms for the transformations of $\mathbf{x}_{Aud}$, $\mathbf{x}_{Vis}$, and $\mathbf{x}_{RF}$.


\noindent \textbf{Step \noindent \tikz[baseline=(char.base)]\node[shape=circle,draw,inner sep=1pt] (char) {\textbf{3}};}: A final fused output $\mathbf{y}$ is computed by combining 
the transformed vectors $\mathbf{y}_1$, $\mathbf{y}_2$, and $\mathbf{y}_3$ based on 
the gating vector $\mathbf{g} = [\,\mathbf{g_1}, \mathbf{g_2}, \mathbf{g_3}\,]^\top$. The combination is performed 
via an element-wise multiplication $\odot$, using $\mathbf{g}$ to weigh each modality's 
contribution:
\vspace{-0.15cm}
\begin{equation}
    \mathbf{y} = \mathbf{g_{1}} \odot \mathbf{y}_1 \;+\; \mathbf{g_{2}} \odot \mathbf{y}_2 
                \;+\; \mathbf{g_{3}} \odot \mathbf{y}_3.
    \vspace{-0.15cm}
\end{equation}
This equation ensures that the unimodal features from audio, video, and RF 
are fused in a controlled manner, based on the learned contribution scores 
stored in the gating vector~$\mathbf{g}$.

\vspace{-0.2cm}
\section{Results and discussion}

\label{Results and discussion}
\subsection{ Experimental Setup}
\subsubsection{Hardware Characteristics}

The training of the unimodal and multi-modal models was performed on a single GPU, the GeForce GTX 1080 Ti, built on the Pascal architecture, featuring 584 CUDA cores, 11 GB of GDDR5X VRAM, and a 352-bit memory interface, offering a memory bandwidth of 484 GB/s. The training settings included the \textsc{Adam} optimizer, \textsc{LRCosineAnnealing} for learning rate scheduling within the range of 0.001 to 0.01, and a budget of 20 epochs. 
After training, the DNNs were run on the Jetson Orin Nano module, powered by a 6-core ARM Cortex-A78AE 64-bit CPU. The system is equipped with 8 GB of 128-bit LPDDR5 memory and a 1024-core NVIDIA Ampere architecture GPU with 32 Tensor Cores.

\vspace{-0.1cm}
\subsubsection{Dataset Overview and Distribution}

This study utilizes a multi-modal dataset integrating audio, video, and RF sensor data, categorized into two classes: drone (class 0) and no drone (class 1). Each data sample is temporally synchronized, comprising a 0.25-second audio segment, a stack of 7 video frames, and a single RF spectrogram. The audio data captures drone flight sounds while differentiating them from general environmental noise. The video data includes drone images at varying distances, along with non-drone objects such as clouds, helicopters, birds, and buildings. The RF data consists of I/Q samples, later transformed into spectrograms to analyze drone-specific communication patterns. A detailed breakdown of the dataset structure, including sample sizes and recording parameters, is provided in Appendix~\ref{appendix B}.
 \vspace{-0.3cm}
\begin{table}[H]\small
\centering
\captionsetup{font=small}
\caption{Detailed data composition across training, validation, and testing phases.}
 \vspace{-0.4cm}
\label{dataset_breakdown}
\resizebox{\columnwidth}{!}{%
\begin{tabular}{l l c c c c c}
\toprule
\textbf{Data type} & \textbf{Class} & \textbf{Files} & \textbf{Audio} & \textbf{Visual} & \textbf{RF} & \textbf{Duration (s)} \\
\midrule
\multirow{2}{*}{\textbf{Train}} & Drone (0) & 118 & 4720 & 33040 & 4720 & \multirow{2}{*}{2120} \\
& No-drone (1) & 94 & 3760 & 26320 & 3760 & \\
\midrule
\multirow{2}{*}{\textbf{Validation}} & Drone (0) & 20 & 800 & 5600 & 800 & \multirow{2}{*}{320} \\
& No-drone (1) & 12 & 480 & 3360 & 480 & \\
\midrule
\multirow{2}{*}{\textbf{Test}} & Drone (0) & 21 & 840 & 5880 & 840 & \multirow{2}{*}{330} \\
& No-drone (1) & 12 & 480 & 3360 & 480 & \\
\midrule
\multicolumn{2}{l}{\textbf{Total}} & \textbf{277} & \textbf{11080} & \textbf{77560} & \textbf{11080} & \textbf{2770} \\
\bottomrule
\end{tabular}
}
\vspace{-0.35cm}
\end{table}

A structured distribution is implemented to ensure robust model development, with 77\% of the dataset allocated for training, 11\% for validation, and 12\% for testing. The training and validation phases rely on real, non-augmented data, allowing models to learn drone detection features without augmentation effects. In contrast, the test dataset includes noise augmentation applied synchronously across all three modalities within each sample, simulating real-world conditions where sensor degradation can affects all data streams simultaneously. The complete dataset composition across training, validation, and testing phases is summarized in Table \ref{dataset_breakdown}.
The trained models are first evaluated on the entire test dataset to establish a baseline performance. For a more granular assessment, the test dataset (1320 samples) is further categorized based on lighting conditions (daylight, sunset) and location types (urban, non-urban), as presented in Table \ref{dataset_breakdown_conditions}.

\begin{table}[H]\small
\centering
\vspace{-0.3cm}
\captionsetup{font=small}
\caption{Test dataset breakdown by environmental conditions.}
\vspace{-0.4cm}
\label{dataset_breakdown_conditions}
\resizebox{\columnwidth}{!}{%
\begin{tabular}{l l c c c c}
\toprule
\multicolumn{2}{c}{\multirow{2}{*}{\textbf{Environmental Condition}}} & \multicolumn{2}{c}{\textbf{Lighting}} & \multicolumn{2}{c}{\textbf{Location}} \\
\cmidrule(lr){3-4} \cmidrule(lr){5-6}
\multicolumn{2}{c}{} & \textbf{Daylight} & \textbf{Sunset} & \textbf{Urban} & \textbf{Non-Urban} \\
\midrule
\multirow{2}{*}{\textbf{Class}} & Drone (0) & 360 & 480 & 160 & 680 \\
& No-drone (1) & 320 & 160 & 200 & 280 \\
\midrule
\multicolumn{2}{l}{\textbf{Total samples per condition}} & \textbf{680} & \textbf{640} & \textbf{360} & \textbf{960} \\
\bottomrule
\end{tabular}
}
\vspace{-0.4cm}
\end{table}

\subsubsection{Evaluation Scenarios}
We evaluated the system's robustness using two distinct scenarios based on the level of noise distortion introduced in the augmented data: scenario 1 (low noise) and scenario 2 (high noise). Noise was added to each data sample from the audio, visual, and RF sensors in the test set, while the model was trained exclusively on real data to avoid biasing the learning process and ensure a fair evaluation.  The augmentation levels were carefully selected to simulate real-world noise conditions, ensuring the test set remains both realistic and representative of practical scenarios.  The distortion in each scenario was quantified using specific metrics: MCD (dB) for audio, SSIM for video, and SNR (dB) for RF data, as described in Section \ref{Data Augumentation} and summarized in Table \ref{tab:scenario_quantification}.

\vspace{-0.25cm}
\begin{table}[H]\small
\centering
\caption{Quantification of augmentation levels for noisy augmented scenarios.}
\vspace{-0.4cm} 
\label{tab:scenario_quantification}
\scalebox{0.9}{
\begin{tabular}{|l|c|c|c|}
\hline
\textbf{Scenario} & \textbf{Audio (MCD) } \cite{saldanha2022data} & \textbf{Visual (SSIM)} \cite{akbiyik2023data} & \textbf{RF (SNR)} \cite{shen2023toward} \\
\hline
1. Low Noise & 3.54 dB & 0.9597 & 15-20 dB \\
2. High Noise & 11.82 dB & 0.7271 & 5-10 dB \\
\hline
\end{tabular}
}
\vspace{-0.3cm}
\end{table}

The low noise scenario represents minimal distortion, where audio, visual, and RF data maintain relatively good quality. In contrast, the high noise scenario reflects significant distortion, simulating harsh real-world conditions with substantial noise and interference across all modalities.

\vspace{-0.2cm}
\subsection{Unimodal Detection Analysis}
This section evaluates the performance of each modality—audio, video, and RF—independently to highlight their strengths and limitations under varying levels of environmental noise and distortion. The evaluation was conducted on real data (non-augmented) and in two noisy augmented scenarios (scenario 1 and scenario 2). 

\begin{table*}[]\small
\centering
\caption{Unimodal model performance under real and augmented noisy data.}
\vspace{-0.4cm} 
\label{tab:unimodal_performance}
\scalebox{0.9}{
\begin{tabular}{@{}ll ccccc ccccc ccccc c@{}}
\toprule
\multicolumn{2}{c}{} &
\multicolumn{5}{c}{\textbf{Real Data}} &
\multicolumn{10}{c}{\textbf{Noisy Augmented Data}} &
\multirow{4}{*}{\textbf{Detec. Time}} \\

\cmidrule(lr){3-7}\cmidrule(lr){8-17}
& & & & & & & 
\multicolumn{5}{c}{\textbf{Scenario 1}} &
\multicolumn{5}{c}{\textbf{Scenario 2}} &
\\
\cmidrule(lr){8-12}\cmidrule(lr){13-17}
& & \textbf{Acc} & \textbf{Prec} & \textbf{Rec} & \textbf{F1} & \textbf{F1-M} &
\textbf{Acc} & \textbf{Prec} & \textbf{Rec} & \textbf{F1} & \textbf{F1-M} &
\textbf{Acc} & \textbf{Prec} & \textbf{Rec} & \textbf{F1} & \textbf{F1-M} &
\\
& & (\%) & (\%) & (\%) & (\%) & (\%) &
(\%) & (\%) & (\%) & (\%) & (\%) &
(\%) & (\%) & (\%) & (\%) & (\%) &
\textbf{(ms)}
\\
\midrule
\multirow{2}{*}{\textbf{Audio}} 
& \textbf{LeNet} 
  & 90.00 & 97.97 & 86.07 & 91.63 & 89.60
  & 82.50 & 87.55 & 84.52 & 86.01 & 81.33
  & 63.64 & 63.64 & 100.00 & 77.78 & 38.89
  & 1.17 \\
& \textbf{VGG-19} 
  & 63.64 & 63.64 & 100.00 & 77.78 & 38.89
  & 63.64 & 63.64 & 100.00 & 77.78 & 38.89
  & 63.64 & 63.64 & 100.00 & 77.78 & 38.89
  & 2.13 \\
\midrule
\multirow{2}{*}{\textbf{Video}} 
& \textbf{ResNet-10}
  & 96.97 & 95.45 & 100.00 & 97.67 & 96.66
  & 74.47 & 71.37 & 100.00 & 83.29 & 64.60
  & 64.02 & 63.88 & 100.00 & 77.96 & 40.01
  & 2.33 \\
& \textbf{MobileNet}
  & 86.14 & 94.69 & 82.86  & 88.38 & 85.60
  & 38.56 & 82.22 & 4.40   & 8.36  & 31.08
  & 37.65 & 100.00 & 2.02 & 3.97  & 28.90
  & 2.11 \\
\midrule
\multirow{2}{*}{\textbf{RF}} 
& \textbf{ResNet-10}
  & 98.41 & 98.69 & 98.81 & 98.75 & 98.28
  & 81.06 & 81.58 & 90.71 & 85.91 & 78.52
  & 62.27 & 65.77 & 84.88 & 74.12 & 52.28
  & 2.42 \\
& \textbf{MobileNet}
  & 92.80 & 93.47 & 95.36 & 94.40 & 92.16
  & 79.09 & 78.60 & 92.26 & 84.88 & 75.49
  & 66.52 & 67.30 & 92.14 & 77.79 & 54.89
  & 1.80 \\
\bottomrule
\end{tabular}
}
\vspace{-0.4cm} 
\end{table*}

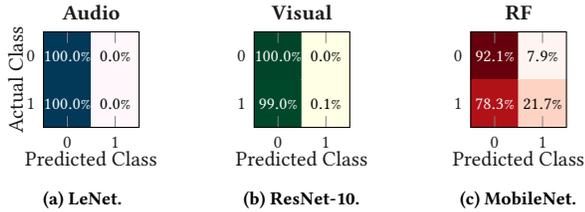
\begin{figure}[h]
    \centering
    \vspace{-0.3cm}
    \setlength\fwidth{0.15\columnwidth}
    \setlength\fheight{0.15\columnwidth}
    \subcaptionbox{\label{lenet_audio}LeNet.}%
    {\input{confusion_matrices/Scenario_2_Unimodal/lenet_audio}}%
    \subcaptionbox{ResNet-10.\label{resnet10_video}}[0.20\textwidth]{%
        \input{confusion_matrices/Scenario_2_Unimodal/resnet10_video}%
    }
    \subcaptionbox{\label{mobilenet_rf}MobileNet.}%
    {\input{confusion_matrices/Scenario_2_Unimodal/mobilenet_rf}}%
    \vspace{-0.4cm}
    \caption{Confusion matrices for the best performing unimodal models in scenario 2.}
    \vspace{-0.25cm}
    \label{confusion_unimodal}
\end{figure}

\noindent \textbf{Audio-based Detection.}~
LeNet and VGG-19 architectures were evaluated for audio-based detection. In real-world data, LeNet outperformed VGG-19, achieving 90.00\% accuracy compared to 63.64\% (Table~\ref{tab:unimodal_performance}). This superior performance is attributed to LeNet’s ability to extract relevant spectral features, resulting in higher precision, recall, and F1-score. It also demonstrated greater efficiency with a 1.17 ms detection time, compared to 2.13 ms for VGG-19. In noisy augmented data, both models showed a decline in accuracy. LeNet remained relatively robust in scenario 1, achieving 82.5\% accuracy, but dropped to 63.64\% in the more challenging scenario 2, revealing its sensitivity to high noise. The confusion matrix for audio-LeNet in scenario 2 (Figure~\ref{lenet_audio}) shows a 100\% True Positive Rate (TPR) but also a 100\% False Positive Rate (FPR), misclassifying all non-drone instances as drones, rendering it ineffective under extreme noise. VGG-19 consistently performed poorly across all conditions, maintaining a 63.64\% accuracy in both scenarios.

\noindent \textbf{Video-based Detection.}~Visual detection performed well in real data, with ResNet-10 (96.97\% accuracy) outperforming MobileNet (86.14\%), as shown in Table~\ref{tab:unimodal_performance}. ResNet-10’s deeper architecture and residual connections helped capture complex spatial features, minimizing missed detections. MobileNet, though more efficient (2.11 ms detection time), had lower recall (82.86\%), making it more prone to missing drones. However, both models struggled in noisy augmented data. ResNet-10’s accuracy dropped to 74.47\% in scenario 1 and 64.02\% in scenario 2, while MobileNet’s fell to 38.56\% and 37.65\%, respectively. The decline highlights the sensitivity of vision-based detection to noise, blur, and occlusions. The confusion matrix for ResNet-10 in scenario 2 (Figure~\ref{resnet10_video}) shows a 100\% TPR, detecting all drones, but a FPR of 99.0\% and a True Negative Rate (TNR) of only 0.1\%, misclassifying nearly all non-drone instances. This pattern mirrors audio-LeNet, indicating a failure to discriminate effectively in highly degraded visual conditions.

\noindent \textbf{RF-based Detection.}~RF-based detection performed well in real data, with ResNet-10 (98.41\% accuracy, 98.75\% F1-score) outperforming MobileNet (92.80\% accuracy, 94.40\% F1-score), as shown in Table~\ref{tab:unimodal_performance}. These results indicate that RF signals provide strong features for drone detection in a relatively clear spectrum. However, performance declined in noisy augmented data. ResNet-10’s accuracy dropped to 81.06\% in scenario 1 and 62.27\% in scenario 2, while MobileNet fell to 79.09\% and 66.52\%, respectively. The confusion matrix for RF-MobileNet (Figure~\ref{mobilenet_rf}) shows a 92.1\% TPR and a 7.9\% False Negative Rate (FNR), detecting most drones, but a high 78.3\% FPR, frequently misclassifying non-drone signals. This highlights the challenge of distinguishing drones from background RF activity in noisy environments.


The instability of individual modalities in noisy conditions highlights the need for multi-modal fusion to achieve more reliable and robust drone detection across diverse environments. In the subsequent multi-modal fusion analysis, scenario 2 is considered, as it presents the most challenging conditions for the detection system.

\begin{table*}[!h]\small
\centering
\caption{Performance metrics for the best performing two-modal fusions under real and noisy augmented data.}
\vspace{-0.4cm} 
\label{tab:fusion_performance_two_modal_best}
\scalebox{0.9}{
\begin{tabular}{@{}llllcccccccccc@{}}
\toprule
\multicolumn{3}{c}{} & \multicolumn{5}{c}{\textbf{Real Data}} & \multicolumn{5}{c}{\textbf{Noisy Augmented Data}} & \multirow{2}{*}{\textbf{Detec. Time}} \\
\cmidrule(lr){4-8} \cmidrule(lr){9-13}
\textbf{Fusion} & \textbf{Modality} & \textbf{Models} & \textbf{Acc} & \textbf{Prec} & \textbf{Rec} & \textbf{F1} & \textbf{F1-M} & \textbf{Acc} & \textbf{Prec} & \textbf{Rec} & \textbf{F1} & \textbf{F1-M} & \multirow{3}{*}{\textbf{(ms)}} \\
\textbf{Type} & \textbf{Combination} & & \textbf{(\%)} & \textbf{(\%)} & \textbf{(\%)} & \textbf{(\%)} & \textbf{(\%)} & \textbf{(\%)} & \textbf{(\%)} & \textbf{(\%)} & \textbf{(\%)} & \textbf{(\%)} & \\
\midrule
\multirow{3.5}{*}{\rotatebox[origin=c]{90}{\textbf{Late}}} & \multirow{1}{*}{\textbf{Audio-Visual}} 
& VGG-19 + ResNet-10 & 96.97 & 95.45 & 100.00 & 97.67 & 96.66 & 66.82 & 66.05 & 98.45 & 79.06 & 49.57 & 5.22 \\
\cmidrule(lr){2-14}
& \multirow{1}{*}{\textbf{Audio-RF}} 
& VGG-19 + ResNet-10 & 96.82 & 100.00 & 95.00 & 97.44 & 96.62 & 73.71 & 85.37 & 70.83 & 77.42 & 72.98 & 4.30 \\
\cmidrule(lr){2-14}
& \multirow{1}{*}{\textbf{Visual-RF}} 
& MobileNet + ResNet-10 & 96.97 & 95.66 & 99.76 & 97.67 & 96.67 & 72.50 & 98.57 & 57.62 & 72.73 & 72.50 & 3.32 \\
\midrule
\multirow{3.5}{*}{\rotatebox[origin=c]{90}{\textbf{GMU}}} & \multirow{1}{*}{\textbf{Audio-Visual}} & LeNet + MobileNet & 99.32 & 98.94 & 100.00 & 99.47 & 99.26 & 55.53 & 60.53 & 86.55 & 71.24 & 36.62 & 2.40 \\
\cmidrule(lr){2-14}
& \multirow{1}{*}{\textbf{Audio-RF}} & VGG-19 + ResNet-10 & 97.12 & 100.00 & 95.48 & 97.69 & 96.94 & 80.53 & 92.31 & 75.71 & 83.19 & 80.03 & 4.25 \\
\cmidrule(lr){2-14}
& \multirow{1}{*}{\textbf{Visual-RF}} & MobileNet + MobileNet & 95.38 & 100.00 & 92.74 & 96.23 & 95.13 & 62.42 & 63.23 & 97.86 & 76.82 & 38.81 & 2.27 \\
\bottomrule
\end{tabular}
\
}
\vspace{-0.4cm}
\end{table*}



\vspace{-0.2cm}
\subsection{Dual-Modal Fusion Analysis}
In this section, we evaluate the performance of dual-modal fusion, using
Late and GMU Fusion techniques to analyze how combining modalities improves detection performance. Table \ref{tab:fusion_performance_two_modal_best} presents the best-performing dual-modal fusion strategies under both real and noisy augmented data, while the full results for all modality combinations are provided in Appendix~\ref{Appendix C1- Dual-Modality}.

\noindent\textbf{Best Performing Combinations.}~In real data, 
as shown in Table \ref{tab:fusion_performance_two_modal_best}, all the listed best combinations exhibit high accuracy, demonstrating the benefit of combining complementary sensor data when environmental factors are benign. Audio-visual (VGG-19 + ResNet-10) with Late Fusion reaches 96.97\% accuracy, highlighting the synergy of visual and auditory cues in real data. Audio-RF and visual-RF fusions also achieve strong accuracies of 96.82\% and 96.97\% respectively, underscoring the effectiveness of dual-modality integration.
 However, in noisy augmented data, dual-modal fusion experiences a noticeable decline in performance, highlighting its sensitivity to high environmental noise.
Visual-RF with Late Fusion shows a noticeable accuracy drop to 72.50\%, indicating vulnerability when both visual and RF data streams are degraded. Audio-visual with Late Fusion also sees a performance decrease, reaching 66.82\% accuracy in noisy environments. In contrast, audio-RF using GMU Fusion demonstrates superior resilience, maintaining an accuracy of 80.53\% under noisy conditions. This highlights audio-RF GMU Fusion as the most robust dual-modal approach in challenging environments.

\begin{figure}[!h]
    \centering
    \vspace{-0.4cm}
    \setlength\fwidth{0.15\columnwidth}
    \setlength\fheight{0.15\columnwidth}
    \subcaptionbox{VGG-19 + ResNet-10.\label{late_audio_visual_vgg_resnet10}}[0.15\textwidth]{%
        \input{confusion_matrices/Dual_Modalities/Late/audio_visual_vgg_resnet10}%
    }
    \subcaptionbox{VGG-19 + ResNet-10.\label{gmu_audio_rf_vgg_resnet10}}[0.15\textwidth]{%
        \input{confusion_matrices/Dual_Modalities/Gmu/audio_rf_vgg_resnet10}%
    }
    \subcaptionbox{MobileNet + ResNet-10.\label{late_visual_rf_mobilenet_resnet10}}[0.16\textwidth]{%
        \input{confusion_matrices/Dual_Modalities/Late/visual_rf_mobilenet_resnet10}%
    }

    \vspace{-0.4cm}
    \caption{Confusion matrices for the best performing dual-modality fusions in noisy augmented data: (blue) Late Fusion; (green) GMU Fusion.}
    \vspace{-0.5cm}
    \label{confusion_matrices_dual_modalities_best}
\end{figure}
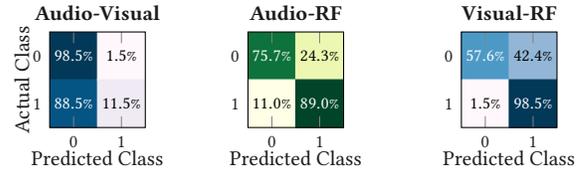

\noindent \textbf{Confusion Matrices.}~Analyzing the confusion matrices for augmented noisy conditions (Figure \ref{confusion_matrices_dual_modalities_best}) provides further insights. For Late Fusion in Figure~\ref{late_audio_visual_vgg_resnet10}, audio-visual (VGG-19 + ResNet-10) shows high drone detection capability (TPR: 98.5\%) but also a high false alarm rate (FPR: 88.5\%), indicating limited ability to reject non-drone instances (TNR: 11.5\%). Visual-RF (MobileNet + ResNet-10) with Late Fusion in Figure~\ref{late_visual_rf_mobilenet_resnet10}, conversely, excels at rejecting non-drones (TNR: 98.5\%, FPR: 1.5\%) but struggles to detect drones effectively (TPR: 57.6\%, FNR: 42.4\%). In contrast, Figure~\ref{gmu_audio_rf_vgg_resnet10}, audio-RF GMU fusion (VGG-19 + ResNet-10) achieves a more balanced performance. It provides a good drone detection rate (TPR: 75.7\%) while maintaining a significantly lower false alarm rate compared to audio-visual Late Fusion (FPR: 11.0\%, TNR: 89.0\%).

\noindent \textbf{Late vs. GMU Fusion.}~
We observe that Late Fusion generally achieves slightly higher accuracy in real data, likely due to its simplicity and effective combination of independently strong unimodal predictions. However, GMU Fusion demonstrates a greater capacity to maintain robustness in augmented noisy environments, particularly evident in the audio-RF combination. This suggests that GMU Fusion's ability to dynamically gate and integrate features at an intermediate level allows it to better adapt to noisy or partially degraded sensor inputs, leading to more reliable detection in challenging scenarios.
Furthermore, it is important to acknowledge the slight increase in detection time when moving from unimodal to dual-modality systems, as seen in Table \ref{tab:fusion_performance_two_modal_best}. For example, the detection time for the best-performing GMU Fusion audio-RF (VGG-19 + ResNet10-10) is 4.25 ms, slightly higher than the unimodal VGG (2.13 ms) or ResNet-10 (2.42 ms). 
Despite these trade-offs, dual-modality approaches still struggle to balance high detection accuracy and low false alarms, motivating the exploration of tri-modality fusion for a more robust drone detection framework.

\vspace{-0.28cm}
\subsection{Proposed Tri-Modal Fusion Analysis }

We evaluate our tri-modal fusion approach, integrating audio, visual, and RF data, focusing on the best-performing combination from Table \ref{tab:fusion_performance_three_modal_best}. An overview of all combinations is in Appendix~\ref{Appendix C2- Tri-Modality}.



\noindent \textbf{Best Performing Combination.}~
The tri-modal fusion approach achieves consistently high performance, matching the near-perfect accuracy of the best dual-modal setups while providing greater stability across different model configurations in real data. As shown in Table \ref{tab:fusion_performance_three_modal_best}, both VGG-19 + MobileNet + ResNet-10 (Late Fusion) and LeNet + ResNet-10 + ResNet-10 (GMU Fusion) maintain accuracy above 96.85\%. This highlights the effectiveness of tri-modal fusion in leveraging the complementary strengths of audio, visual, and RF data when environmental factors are not disruptive. However, the true advantage of tri-modal fusion emerges in augmented noisy conditions, where it demonstrates significantly enhanced robustness compared to unimodal and dual-modal systems. As shown in Table \ref{tab:fusion_performance_three_modal_best}, the best-performing tri-modal model, VGG-19 + MobileNet + ResNet-10 (Late Fusion), achieves 83.26\% accuracy and an 85.16\% F1-score, outperforming the best dual-modality configuration (audio-RF GMU Fusion) at 80.53\% accuracy (Table \ref{tab:fusion_performance_two_modal_best}). This highlights the crucial role of a third modality in improving system resilience, enabling the model to compensate for sensor degradation and maintain reliable detection despite environmental interference.

\noindent \textbf{Confusion Matrices.}~The confusion matrices in Figure \ref{confusion_matrices_three_modalities} provide a visual perspective on the performance improvements.
For Late Fusion (Figure \ref{late_3_modal_lenet_mobilenet_resnet10}), VGG-19 + MobileNet + ResNet-10 achieves high TNR (96.9\%), ensuring strong background rejection, and a low FPR (3.1\%), maintaining balanced classification. Compared to the best dual-modal fusion (Figure \ref{gmu_audio_rf_vgg_resnet10}), it reduces false positives by nearly 7.9\% (FPR: 3.1\% vs. 11.0\%), improves background rejection (TNR: 96.9\% vs. 89.0\%), and maintains recall (TPR: 75.5\% vs. 75.7\%). These improvements make tri-modal fusion more precise and stable, ensuring fewer misclassifications while preserving strong drone detection, making it more reliable in complex environments.
For GMU Fusion, LeNet + ResNet-10 + ResNet-10 (Figure \ref{gmu_3_modal_lenet_resnet10_resnet10}) ensures 100\% TPR, detecting all drones with no missed instances (FNR: 0\%). However, this comes at the cost of a high false alarm rate (FPR: 67.3\%) and weak background rejection (TNR: 32.7\%), leading to frequent misclassification of non-drone signals as drones.

\begin{figure}[!h]
    \centering
    \vspace{-0.3cm}
    \setlength\fwidth{0.15\columnwidth}
    \setlength\fheight{0.15\columnwidth}
    \subcaptionbox{\label{late_3_modal_lenet_mobilenet_resnet10}VGG-19 + MobileNet + ResNet-10.}[0.23\textwidth]%
    {\input{confusion_matrices/late_3_modal/lenet_mobilenet_resnet10}}%
    \subcaptionbox{\label{gmu_3_modal_lenet_resnet10_resnet10}LeNet + ResNet-10 + ResNet-10.}[0.23\textwidth]%
    {\input{confusion_matrices/gmu_3_modal/lenet_resnet10_resnet10}}%
    \vspace{-0.4cm}
    \caption{Confusion matrices for tri-modality fusions in noisy augmented data: (blue) Late Fusion; (green) GMU Fusion.}
    \vspace{-0.31cm}
    \label{confusion_matrices_three_modalities}
\end{figure}

\begin{table*}[!ht]\small
\centering
\caption{Performance metrics for the best performing three-modal fusions under real and noisy augmented data.}
\vspace{-0.4cm}
\label{tab:fusion_performance_three_modal_best}
\begin{adjustbox}{width=\textwidth,center}
\begin{tabular}{@{}llllccccccccccc@{}}
\toprule
\multicolumn{3}{c}{} & \multicolumn{5}{c}{\textbf{Real Data}} & \multicolumn{5}{c}{\textbf{Noisy Augmented Data}} & \multirow{2}{*}{\textbf{Detec. Time}} & \multirow{2}{*}{\textbf{Ener. Cons.}}  \\
\cmidrule(lr){4-8} \cmidrule(lr){9-13}
\textbf{Fusion} & \textbf{Modality} & \textbf{Models} & \textbf{Acc} & \textbf{Prec} & \textbf{Rec} & \textbf{F1} & \textbf{F1-M} & \textbf{Acc} & \textbf{Prec} & \textbf{Rec} & \textbf{F1} & \textbf{F1-M} & \\
\textbf{Type} & \textbf{Combination} & & \textbf{(\%)} & \textbf{(\%)} & \textbf{(\%)} & \textbf{(\%)} & \textbf{(\%)} & \textbf{(\%)} & \textbf{(\%)} & \textbf{(\%)} & \textbf{(\%)} & \textbf{(\%)} & \textbf{(ms)} & \textbf{(mJ)} \\
\midrule
\multirow{1}{*}{{\textbf{Late}}} & \multirow{1}{*}{\textbf{Audio-Visual-RF}} 
&  VGG-19 + MobileNet + ResNet-10 & 96.89 & 95.35 &100.00 & 97.62 & 96.58 & 83.26 & 97.69 & 75.48 & 85.16 & 82.98 & 6.09 & 75.27  \\
\midrule
\multirow{1}{*}{{\textbf{GMU}}} & \multirow{1}{*}{\textbf{Audio-Visual-RF}} 
& LeNet + ResNet-10 + ResNet-10 & 96.97 & 95.45 & 100.00 & 97.67 & 96.66 & 75.53 & 72.23 & 100.00 & 83.87 & 66.58 & 6.48 & 80.09\\
\bottomrule
\end{tabular}
\end{adjustbox}
\vspace{-0.4cm}
\end{table*}

\noindent \textbf{Late vs. GMU Fusion.}~In contrast to dual-modality fusion where GMU Fusion showed robustness advantages in noisy augmented data, in tri-modality, Late Fusion emerges as the superior technique in terms of overall accuracy and balanced performance. Late Fusion’s approach of combining decisions from independently optimized unimodal models appears to be particularly effective when integrating three diverse sensor streams, providing both high accuracy and robustness. 
Despite its added complexity, tri-modal fusion remains efficient, achieving detection times of 5–6 ms (Table \ref{tab:fusion_performance_three_modal_best}), only slightly higher than dual-modal fusion while staying well within real-time constraints for drone detection and response. This efficiency extends to energy consumption, with the best combination in Late Fusion consuming 75.27 mJ and GMU Fusion requiring 80.09 mJ. These fusion approaches were tested on the Jetson Orin Nano using the GPU, demonstrating their feasibility for deployment on resource-constrained devices.




\noindent \textbf{Light and Environmental Conditions Effects.}  
In the following, we evaluate the impact of lighting and environmental conditions on the performance of the proposed tri-modal fusion approach.

\begin{figure}[!h]
    \centering
    \vspace{-0.3cm}
    \setlength\fwidth{0.15\columnwidth}
    \setlength\fheight{0.15\columnwidth}
    \subcaptionbox{\label{day_late_VGG + MobileNet + ResNet10}Day.}%
    {\input{confusion_matrices/Day/late_VGG_+_MobileNet_+_ResNet10}}%
    \subcaptionbox{\label{Sunset_Late_VGG + MobileNet + ResNet10}Sunset.}%
    {\input{confusion_matrices/Sunset/Late_VGG_+_MobileNet_+_ResNet10}}%
    \subcaptionbox{\label{day_gmu_lenet + Resnet10 + ResNet10}Day.}%
    {\input{confusion_matrices/Day/gmu_lenet_+_Resnet10_+_ResNet10}}%
    \subcaptionbox{\label{Sunset_GMU_LeNet + ResNet10 + ResNet10}Sunset.}%
    {\input{confusion_matrices/Sunset/GMU_LeNet_+_ResNet10_+_ResNet10}}%
    \vspace{-0.4cm}
    \caption{Confusion matrices showing the impact of lighting conditions across three modalities for Late (blue) and GMU Fusion (green).}
    \vspace{-0.35cm}
    \label{light_effect_three_modalities}
\end{figure}
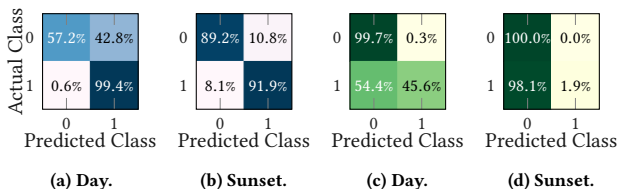

\noindent \tikz[baseline=(char.base)]\node[shape=circle,draw,inner sep=1pt] (char) {1}; \noindent \textit{Light Condition Analysis.}~The impact of lighting on the performance of the proposed tri-modal fusion approach was evaluated under noisy augmented data (scenario 2) during both \textit{daytime} and \textit{sunset}. The results in Table \ref{three_modalities_ligh_locations} and the confusion matrices in Figure \ref{light_effect_three_modalities} provide deeper insight into how illumination variations influence detection performance. The sunset conditions generally provide more favorable and stable illumination, leading to improved performance across all fusion strategies. Under \textbf{\textit{Late Fusion}}, the VGG-19 + MobileNet + ResNet-10 combination performed best at sunset, achieving 89.84\% accuracy, with strong background rejection (TNR: 91.9\%) and a low false alarm rate (FPR: 8.1\%). The consistent lighting conditions at sunset help reduce visual challenges, such as glare and high-contrast shadows, leading to higher recall (TPR: 89.2\%), fewer missed detections (FNR: 10.8\%), and a high F1-score (92.94\%). During daytime, accuracy dropped to 77.06\%, with a significant decline in drone detection (TPR: 57.2\%) and increased missed detections (FNR: 42.8\%). However, background rejection remained strong (TNR: 99.4\%), minimizing false alarms. The fluctuating light intensity and shadows degraded video-based detection, but audio and RF data helped compensate, ensuring more consistent overall performance. This highlights the effectiveness of multi-sensor fusion in mitigating modality-specific weaknesses, making the system more resilient to environmental variations.
\textbf{\textit{GMU Fusion}} maintains high recall across lighting conditions but struggles with false positives. During daytime, it achieves 74.26\% accuracy, detecting almost all drones (TPR: 99.7\%) with minimal missed detections (FNR: 0.3\%). However, background rejection is weak (TNR: 45.6\%), leading to a high false alarm rate (FPR: 54.4\%). At sunset, accuracy improves slightly to 75.47\%, with perfect drone detection (TPR: 100\%) and no missed detections (FNR: 0\%). However, background rejection deteriorates significantly (TNR: 1.9\%), resulting in an extremely high false alarm rate (FPR: 98.1\%), making it unreliable in environments that require precise classification.

\begin{table*}[!ht]
\small
\centering
\caption{Performance metrics for the top three-modal fusions combos across lighting (day/sunset) and location (urban/non-urban).}
\vspace{-0.4cm}
\label{three_modalities_ligh_locations}
\scalebox{0.74}{
\begin{tabular}{@{}ll ccccc ccccc ccccc ccccc@{}}
\toprule
\multicolumn{2}{c}{\textbf{Fusion}} 
& \multicolumn{10}{c}{\textbf{Lighting Condition}} 
& \multicolumn{10}{c}{\textbf{Location}}
\\
\cmidrule(lr){1-2}\cmidrule(lr){3-12}\cmidrule(lr){13-22}

\textbf{Type} & \textbf{Models Combination}
& \multicolumn{5}{c}{\textbf{Day}}
& \multicolumn{5}{c}{\textbf{Sunset}}
& \multicolumn{5}{c}{\textbf{Urban}}
& \multicolumn{5}{c}{\textbf{Non-Urban}}
\\
\cmidrule(lr){3-7}\cmidrule(lr){8-12}\cmidrule(lr){13-17}\cmidrule(lr){18-22}

& &
\textbf{Acc} & \textbf{Prec} & \textbf{Rec} & \textbf{F1} & \textbf{F1-M}
& \textbf{Acc} & \textbf{Prec} & \textbf{Rec} & \textbf{F1} & \textbf{F1-M}
& \textbf{Acc} & \textbf{Prec} & \textbf{Rec} & \textbf{F1} & \textbf{F1-M}
& \textbf{Acc} & \textbf{Prec} & \textbf{Rec} & \textbf{F1} & \textbf{F1-M}
\\

& &
(\%) & (\%) & (\%) & (\%) & (\%)
& (\%) & (\%) & (\%) & (\%) & (\%)
& (\%) & (\%) & (\%) & (\%) & (\%)
& (\%) & (\%) & (\%) & (\%) & (\%)
\\
\midrule

\multirow{1}{*}{Late} 
  & VGG-19 + MobileNet + ResNet-10
    & 77.06 & 99.04 & 57.22 & 72.54 & 76.42
    & 89.84 & 97.05 & 89.17 & 92.94 & 87.42
    & 78.89 & 87.50 & 61.25 & 72.06 & 77.55
    & 84.90 & 99.81 & 78.82 & 88.09 & 83.73
\\
\midrule

\multirow{1}{*}{GMU} 
  & LeNet + ResNet-10 + ResNet-10
    & 74.26 & 67.35 & 99.72 & 80.40 & 71.46
    & 75.47 & 75.35 & 100.00 & 85.94 & 44.81
    & 46.39 & 45.33 & 100.00 & 62.38 & 34.57
    & 86.35 & 83.85 & 100.00 & 91.21 & 80.34
\\
\bottomrule
\end{tabular}
}
\vspace{-0.4cm}
\end{table*}

\begin{figure}[!h]
    \centering
    \vspace{-0.3cm}
    \setlength\fwidth{0.15\columnwidth}
    \setlength\fheight{0.15\columnwidth}
    \subcaptionbox{\label{urban_late_VGG + MobileNet + ResNet10}Urban.}%
    {\input{confusion_matrices/Urban/late_VGG_+_MobileNet_+_ResNet10}}%
    \subcaptionbox{\label{non_urban_late_VGG + MobileNet + ResNet10}Non-Urban.}%
    {\input{confusion_matrices/Non_Urban/late_VGG_+_MobileNet_+_ResNet10.tex}}%
    \subcaptionbox{\label{urban_gmu_lenet + Resnet10 + ResNet10}Urban.}%
    {\input{confusion_matrices/Urban/gmu_lenet_+_Resnet10_+_ResNet10}}%
    \subcaptionbox{\label{non_urban_gmu_lenet + Resnet10 + ResNet10}Non-Urban.}%
    {\input{confusion_matrices/Non_Urban/gmu_lenet_+_Resnet10_+_ResNet10}}%
    \vspace{-0.4cm}
    \caption{Confusion matrices illustrating the impact of locations across three modalities for Late (blue) and GMU Fusion (green).}
    \vspace{-0.3cm}
    \label{location_effect_three_modalities}
\end{figure}
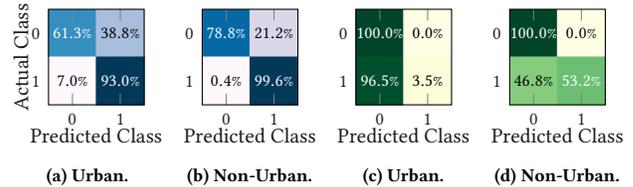

\noindent \tikz[baseline=(char.base)]\node[shape=circle,draw,inner sep=1pt] (char) {2}; \noindent \textit{Location-Based Analysis.}
The results in Table \ref{three_modalities_ligh_locations} and Figure \ref{location_effect_three_modalities} show how urban and non-urban environments affect the tri-modal fusion approach under noisy augmented data (scenario 2). Non-urban settings offer higher accuracy and fewer false alarms across all fusion strategies, benefiting from cleaner RF conditions, clearer audio signals, and a more uniform visual background. Specifically, under \textbf{\textit{Late Fusion}}, detection is significantly stronger in non-urban environments, achieving a higher accuracy (84.90\%), supported by strong background rejection (TNR: 99.6\%) and a minimal false alarm rate (FPR: 0.4\%). Drones are detected more reliably, with a higher TPR (78.8\%) and fewer missed detections (FNR: 21.2\%).
In contrast, urban environments introduce more noise and interference, leading to a drop in accuracy (78.89\%) and a decrease in drone detection (TPR: 61.3\%), with more missed instances (FNR: 38.8\%). However, background rejection remains strong (TNR: 93.0\%), ensuring that non-drone samples are still well classified despite the increased complexity of the scene.  \textbf{\textit{GMU Fusion}}, while maintaining perfect drone detection (TPR: 100\%) in both environments, suffers from extremely high false alarms. In urban areas, performance drops drastically (46.39\%) due to severe background misclassification (TNR: 3.5\%) and an overwhelming false alarm rate (FPR: 96.5\%). While every drone is detected (TPR: 100\%, FNR: 0\%), the model struggles to differentiate drones from background clutter. In non-urban areas, it achieves an accuracy of 86.35\%, benefiting from better background separation (TNR: 53.2\%) and a lower false alarm rate (FPR: 46.8\%) compared to urban environments. These results highlight the trade-off between fusion strategies: Late Fusion offers a more balanced approach, maintaining strong detection performance and minimizing false alarms, making it better suited for urban settings. In contrast, GMU Fusion prioritizes recall, making it more effective in non-urban environments where false positives are less critical.

\vspace{-0.3cm}
\section{Limitations and Discussion}

\noindent \textbf{Dataset Expansion.}~Our dataset focuses on capturing synchronized audio, visual, and RF data in diverse real-world environments, providing a strong basis for multi-modal drone detection. It emphasizes small UAVs, such as the DJI Mini series, and reflects common operational conditions. However, including other types of drones, particularly larger models with different acoustic and RF characteristics, along with additional environmental factors such as rain, fog, and dense vegetation, may further enhance generalizability. Despite this, the dataset remains a valuable resource, offering truly synchronized multi-modal data in uncontrolled settings and serving as a strong benchmark for small UAV detection.

\noindent \textbf{Adversarial Robustness Considerations.}~
While our data augmentation enhances \IW's resilience to natural noise, intentional adversarial attacks by a knowledgeable adversary remain a challenge.  An attacker aware of \IW's tri-modal fusion could strategically manipulate sensor reliability to exploit the system.  For example, subtle visual camouflage, targeted audio masking mimicking background sounds, or intermittent RF interference designed to resemble legitimate signals could be employed. These modality-specific manipulations aim to degrade individual sensor reliability, causing \IW to misweigh sensor inputs within the fusion process and increasing wrong detections or false alarms.  Defending against such intelligent adversarial exploitation, which targets sensor vulnerabilities to perturb the fusion mechanism, is crucial for future work, requiring techniques to enhance sensor-level robustness and develop more resilient fusion strategies.


\noindent \textbf{Counter-UAV Measures.}~This study establishes a strong foundation for drone detection, a key element of counter-UAV systems. Integrating detection with localization, tracking, and mitigation could significantly improve response time and situational awareness, allowing for more effective countermeasures. While these aspects fall outside the scope of this work, a highly reliable detection system remains a critical first step toward a comprehensive counter-UAV strategy.

\vspace{-0.2cm}
\section{Related Work}
\label{Related Work}

\noindent \textbf{Dataset Diversity.}~Many existing drone detection datasets consist of synthetic data or are collected in controlled environments, such as laboratories, test sites, or airport runways, where sensor-to-target visibility is high, background noise is minimal, and RF interference is controlled \cite{svanstrom2021dataset, lenhard2024syndronevision, basak2021combined}. These conditions fail to reflect real-world challenges, where urban noise, lighting variations, occlusions, and overlapping RF signals significantly impact detection performance \cite{bentamou2023real, xu2025radio}. 
In contrast, \IW introduces a novel dataset collected in real-world, uncontrolled environments, covering both urban and non-urban locations under varying lighting conditions (daylight and sunset). This deliberate focus on real-world data acquisition, with synchronized multi-modal sensor streams, enhances the generalizability of our \IW framework.

\noindent \textbf{Data Augmentation and Testing.}~
Data augmentation is crucial for enhancing the robustness of drone detection models. Existing methods typically apply augmentation to only one sensor at a time, such as adding noise to audio or applying visual transformations independently \cite{alla2024sound, lee2023cnn}. However, this approach fails to capture synchronized multi-modal degradations, where real-world factors simultaneously impact multiple sensors, leading to compromised detection accuracy.
Additionally, some approaches train models on synthetically augmented data but evaluate them on non-augmented or minimally augmented test sets \cite{akyon2021track, wisniewski2024drone}. This mismatch creates a performance gap, as models trained under these conditions often struggle with severe real-world noise, making them unreliable for practical deployment \cite{dieter2023quantifying}.
Unlike existing approaches, \IW applies synchronized data augmentation across audio, visual, and RF modalities \textit{exclusively during testing}. This approach enables a rigorous evaluation of system performance under varying levels of multi-modal noise (scenario 1 and scenario 2), effectively measuring the lower bound of accuracy when all sensors experience simultaneous degradation. Importantly, \IW models are trained solely on \textit{ real, non-augmented data}, ensuring that robustness improvements are not a result of overfitting to specific synthetic noise patterns but rather reflect the true effectiveness of multi-modal fusion in handling unseen, real-world degradations. 

\noindent \textbf{Multi-Modality Strategy.}~Early UAV detection relied on single-modality systems, which are highly sensitive to environmental noise, lighting conditions, and RF interference \cite{shi2018hidden,anwar2019machine,ezuma2019detection,medaiyese2021semi}. Dual-modality approaches improve resilience by combining two complementary sensors, but they still fail when both modalities are simultaneously degraded \cite{lee2023cnn,jovanoska2021passive,svanstrom2022drone}. \IW advances the state-of-the-art with a tri-modal (audio-visual-RF) fusion framework, leveraging the strengths of all three modalities to enhance detection robustness in  challenging conditions. 
Unlike existing systems that utilize simpler fusion techniques, \IW incorporates sophisticated architectures that dynamically learn the optimal contribution of each modality during the training process. Approaches relying on basic OR-function logic for detection \cite{shi2018anti, dong2023drone}, equally weighted modalities \cite{svanstrom2021real}, multinomial logistic regression for fixed probability combination \cite{lee2023cnn}, or static ensemble stacking \cite{mccoy2024optimized} inherently lack the adaptability of \IW. These simpler methods are unable to effectively account for the varying reliability of each sensor under different environmental conditions. 
In contrast, \IW's learned fusion weights enable intelligent, context-aware sensor integration, significantly enhancing performance in real-world scenarios.

\noindent \textbf{Real-Time and Energy-Efficiency.}~Real-time drone detection is crucial for timely responses in security and surveillance applications.  Many existing systems, while achieving high accuracy, may not fully address the computational constraints of real-time deployment, or lack detailed reporting on detection times \cite{sricharan2023real,allahham2020deep,medaiyese2021machine,mo2022deep}. Some works focus on optimizing models for edge devices, but may compromise accuracy for efficiency \cite{song2025ednet}. \IW prioritizes both accuracy and real-time feasibility, with its best-performing Late Fusion model achieving a detection time of 6.09 ms per segment, ensuring suitability for real-time applications. Energy efficiency was evaluated on Jetson Orin Nano, measuring 75.27 mJ for Late Fusion and 80.09 mJ for GMU Fusion, demonstrating its capability for edge deployment. This detailed assessment of inference time and energy consumption, often overlooked in related work, highlights \IW's practical deployability for real-world drone detection.

\vspace{-0.2cm}
\section{Conclusions} 
\label{conclusion}


This paper shows that robust drone detection in real-world environments requires multi-sensor data fusion. We present \IW, a tri-modal framework for real-time UAV detection that: (i) achieves 96.89\% accuracy in real data and 83.26\% in noisy augmented data, outperforming unimodal and dual-modal baselines; (ii) remains robust under real-world noise scenarios through targeted data augmentation; (iii) operates in real-time (6.09 ms detection) with low energy consumption (75.27 mJ per detection), making it suitable for resource-constrained devices; and (iv)
generalizes well to diverse urban and non-urban environments under varying lighting conditions, ensuring reliable performance under adverse situations.

\begin{acks}
This research was made possible with the support of the Horizon Europe research and innovation programme of the European Union, under grant agreement number 101092912 (project MLSysOps). 
The authors also gratefully acknowledge the Intelligent and Autonomous Systems Lab (IASL) at the University of California, Irvine (UCI), particularly Sharon Ladron De Guevara Contreras and Marco Levorato for their support and collaboration during the data collection process.

\end{acks}

\bibliographystyle{ACM-Reference-Format}
\bibliography{sample}

\appendix

\section{Technical Specifications of Drones and Sensing Equipment}
\label{appendix A}

\begin{figure}[htbp]
    \centering
        \includegraphics[scale=0.2]{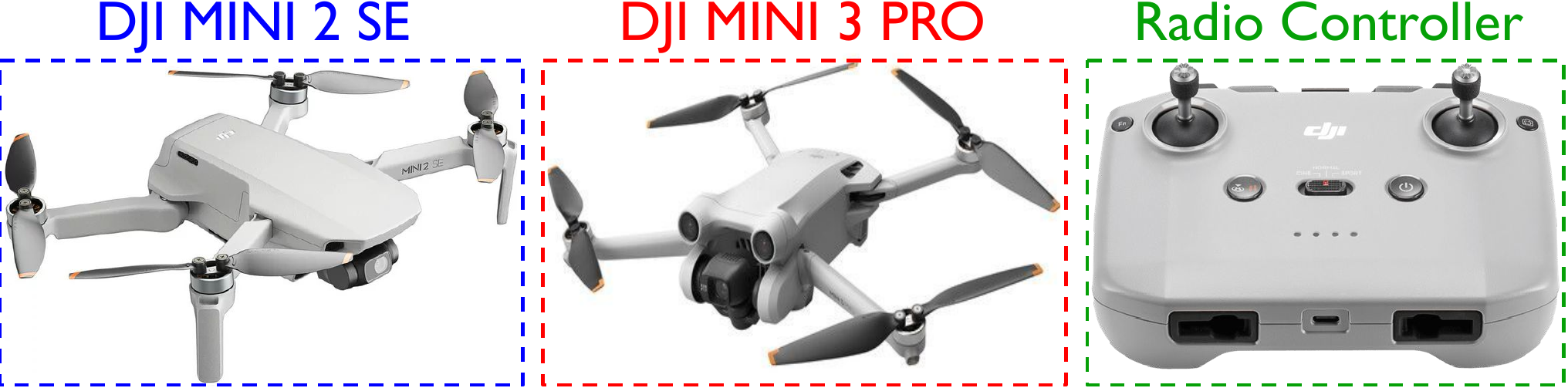}
        \vspace{-0.4cm} 
    \caption{UAVs and their radio controller used during experiments.}
    \label{fig:drones_in_study}
\end{figure}

This appendix provides detailed technical specifications for the drones (Figure~\ref{fig:drones_in_study}) used in our experiments and the sensing equipment employed for data collection. This information is intended to offer a comprehensive understanding of the experimental setup and the technical capabilities of the key components used in this study.
\vspace{-0.22cm} 
\vspace{-0.22cm} 

\noindent \tikz[baseline=(char.base)]\node[shape=circle,draw,inner sep=1pt] (char) {1}; \textbf{Drone Specifications.}
We utilized two commercially available drones, the DJI Mini 2 and DJI Mini 3 Pro, to evaluate our \IW system. These models were selected due to their subtle yet important differences in characteristics such as size, acoustics, and operational capabilities, which pose varying challenges for detection systems. Table \ref{tab:drone-specs} presents a detailed comparison of their key specifications, highlighting the nuances that informed our experimental design.
\begin{table}[h]
\small
\caption{Technical characteristics of drones used in experiments. 
\cite{dji2023mini2se,dji2023minipro}.}
\vspace{-0.35cm} 
\label{tab:drone-specs}
\scalebox{0.9}{
\begin{tabular}{|p{2.5cm}|p{2.2cm}|p{2.2cm}|}
\hline
\textbf{Parameter} & \textbf{DJI Mini 2} & \textbf{DJI Mini 3 Pro} \\
\hline
Size [mm] & 138×81×58 & 145×90×62 \\
\hline
Weight [g] & \multicolumn{2}{c|}{249} \\
\hline
Material & \multicolumn{2}{p{4.8cm}|}{Plastic and metal components; non-reflective and discernible surfaces} \\
\hline
Sound Propeller [dB/m] & 74 & 72 \\
\hline
Diagonal Length [mm] & 213 & 247 \\
\hline
Operating Frequency [GHz] & \multicolumn{2}{c|}{2.400-2.4835, 5.725-5.850} \\
\hline
Max. Transmission Distance [km] & 10 (FCC), 6 (CE), 6 (SRRC), 6 (MIC) & 12 (FCC), 8 (CE), 8 (SRRC), 8 (MIC) \\
\hline
Max Speed [m/s] & \multicolumn{2}{c|}{16 (S Mode), 10 (N Mode), 6 (C Mode)} \\
\hline
\end{tabular}
}
\vspace{-0.35cm} 
\end{table}

As shown in Table \ref{tab:drone-specs}, both drones share a lightweight design at 249g, constructed from plastic and metal components with non-reflective surfaces to minimize visual detectability.  Subtle differences exist in their physical dimensions and acoustic profiles, with the DJI Mini 3 Pro being slightly larger and quieter (72 dB/m) than the DJI Mini 2 (74 dB/m). Both operate in the common 2.4 GHz and 5.8 GHz frequency bands and exhibit similar maximum speeds, but the DJI Mini 3 Pro offers an extended maximum transmission distance. These variations are crucial for evaluating the \IW's ability to discern drones with differing acoustic and signal characteristics.

\begin{figure}[htbp]
    \centering
   \includegraphics[width=0.3\textwidth]{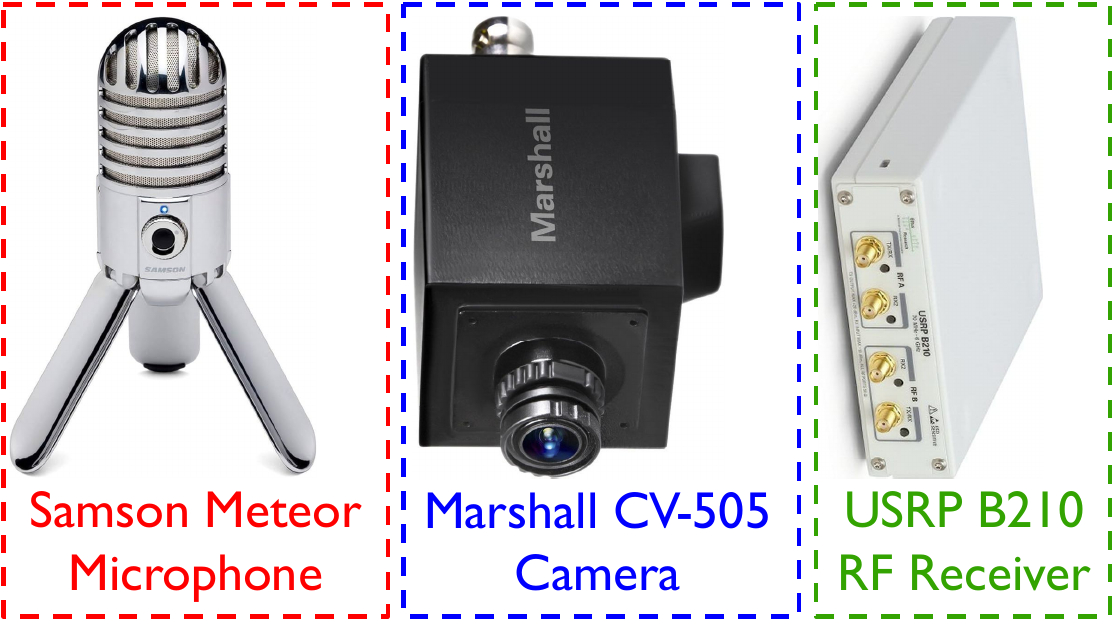}
   \vspace{-0.4cm} 
    \caption{Sensing equipment used during the experiments.}
     \label{fig: sensing equipments}
     \vspace{-0.32cm} 
    \end{figure}

\noindent \tikz[baseline=(char.base)]\node[shape=circle,draw,inner sep=1pt] (char) {2}; \textbf{Sensing Equipment Specifications.}
Our data collection system is designed to capture synchronized audio, visual, and RF data streams, utilizing a suite of sensors. Figure \ref{fig: sensing equipments} illustrates the primary sensing equipment setup, while the subsequent tables detail the technical specifications of each sensor component.
\vspace{-0.2cm} 

\noindent $\bullet$ \textbf{Audio Sensor.}~We employed the Samson Meteor condenser microphone for audio data acquisition, chosen for its high sensitivity and broad frequency response, crucial for capturing the subtle acoustic signatures of small drones. Table \ref{table:samson_meteor_specs} outlines the key specifications of this microphone.

\vspace{-0.3cm} 
\begin{table}[H]
\small
\centering
\caption{Specifications of the Samson Meteor microphone \cite{samsontech2024meteormic}.}
\vspace{-0.4cm} 
\scalebox{0.9}{
\begin{tabular}{l|l}
\hline
\textbf{Parameter}             & \textbf{Value}                                  \\ \hline
Type                           & Condenser                                       \\ 
Polar Pattern                  & Cardioid                                        \\ 
Frequency Response             & 20Hz - 20kHz                                    \\ 
Sample Rate                    & 44.1/48kHz                                      \\ 
Bit Rate                       & 16-bit                                          \\ 
Sensitivity                    & -33 dB/Pa                                       \\ 
                    \hline
\end{tabular}
}
\label{table:samson_meteor_specs}
\vspace{-0.3cm} 
\end{table}

The Samson Meteor is a condenser-type microphone with a cardioid polar pattern, effectively capturing sound sources directly in front while minimizing background noise. Its wide frequency response (20Hz-20kHz) ensures comprehensive capture of the drone's acoustic spectrum. It records high-quality audio at sample rates of 44.1 kHz or 48 kHz with a 16-bit bit depth, and features a sensitivity of -33 dB/Pa, enabling it to detect faint drone sounds even in moderately noisy environments.


\vspace{-0.25cm} 
\begin{table}[h]
\centering
\small
\caption{Specifications of the Marshall CV-505 camera \cite{marshall2024cv505mb}.}
\vspace{-0.4cm} 
\label{table:camera-specs}
\scalebox{0.9}{
\begin{tabular}{l|l}
\textbf{Parameter} & \textbf{Value} \\ \hline
Image Sensor & 2.5 Megapixel 1/3-inch CMOS \\ 
Lens & 3.7mm \\ 
Effective Pixel & 1920(H) x 1080(V) \\ 
Min. Illumination & \begin{tabular}[c]{@{}l@{}}0.2 Lux (Color),\\ 0.1 Lux (Black/White),\\ 0.005 Lux (DSS on)\end{tabular} \\ 
Power Supply & 12V \\ 
Special Features & \begin{tabular}[c]{@{}l@{}}WDR function,
\\ RS-485 remote control\end{tabular} \\ 
Operation Temperature & -10°C $\sim$ 50°C \\ 
\hline
\end{tabular}
}
\vspace{-0.15cm} 
\end{table}
\noindent $\bullet$ \textbf{Video Sensor.}
Visual data was captured using the Marshall CV-505 camera, selected for its compact form factor and high-resolution imaging capabilities, suitable for drone detection in diverse lighting conditions. The detailed specifications of the Marshall CV-505 camera are provided in Table \ref{table:camera-specs}.

The Marshall CV-505 camera utilizes a 2.5 Megapixel 1/3-inch CMOS sensor and a 3.7mm lens, capturing high-resolution video with 1920x1080 effective pixels. Its minimum illumination sensitivity, as low as 0.005 Lux in DSS mode, ensures effective video capture even in low-light conditions.  The camera also features wide dynamic range (WDR) to handle scenes with high contrast lighting and RS-485 remote control for flexible operation.

\noindent $\bullet$ \textbf{RF Sensor.}
We utilized the Universal Software Radio Peripheral (USRP) B210 for RF signal acquisition, a software-defined radio known for its wide frequency range and high sampling rate, essential for capturing drone communication signals. Table \ref{table:usrp-specs} details the specifications of the USRP B210.

\begin{table}[h]
\small
\centering
\vspace{-0.25cm} 
\caption{Specifications of the USRP B210 RF receiver.}
\vspace{-0.4cm} 
\label{table:usrp-specs}
\scalebox{0.9}{
\begin{tabular}{l|l}
\hline
\textbf{Parameter} & \textbf{Value} \\ \hline
Frequency Range & 70 MHz - 6 GHz \\ 
Instantaneous Bandwidth & $\sim$ 56 MHz \\ 
Maximum I/Q sample rate & 61.44 MS/s \\
Number of channels & 2 \\ 
Maximum Receive Gain & 76 dB \\ 
Maximum Transmit Gain & 89.8 dB \\ 
ADC resolution & 12 bits \\ 
\hline
\end{tabular}
}
\vspace{-0.15cm} 
\end{table}

The USRP B210 offers a wide frequency range from 70 MHz to 6 GHz, covering typical drone communication bands. It supports an instantaneous bandwidth of approximately 56 MHz and a maximum I/Q sample rate of 61.44 MS/s, enabling high-fidelity capture of RF signals. With 2 channels and significant gain in both receive (76 dB) and transmit (89.8 dB), the USRP B210 is well-suited for detecting weak drone signals amidst background RF noise. The 12-bit ADC resolution ensures accurate digitization of the received RF signals for subsequent processing and analysis.



\vspace{-0.2cm} 
\section{Dataset Overview}
\label{appendix B}

The dataset collected for this study includes synchronized data from audio, video, and RF sensors, divided into two classes: \textit{drone} and \textit{no drone}. Each data sample consists of ten-second clips, captured within a range of 1 to 50 meters. 

\noindent \textbf{Audio Dataset.}
The audio dataset comprises 277 ten-second clips recorded at a sample rate of 44.1 kHz. 
The \textit{drone} class captures the characteristic sounds of drone flight, while the \textit{no drone} class specifically features general outdoor background sounds collected at the acquisition sites. These include noises from vehicles, voices, and other ambient sounds, providing a robust baseline for distinguishing drone-related noises from environmental sounds. 

\noindent \textbf{Video Dataset.} 
The video dataset consists of 
83,100 frames at a resolution of 640 $\times$ 640 pixels. The \textit{drone} class captures drone activity at distances up to 50 meters, while the \textit{no drone} class includes images of non-drone objects such as clouds, helicopters, birds, buildings, and trees. This variety ensures the system can accurately differentiate drones from other airborne objects and background elements.

\noindent \textbf{RF Dataset.} 
The RF dataset contains I/Q samples recorded at a sampling rate of 55 MS/s in the 2.4 GHz ISM band. These samples were saved in binary files (277 files) and later processed into spectrograms using the STFT, resulting in 11,080 images with a resolution of 640 × 640 pixels.  
The \textit{drone} class in the RF dataset captures the characteristic spectral signatures of drone communication signals.  The \textit{no drone} class encompasses a diverse range of background RF noise and interference from other signal sources. 

\vspace{-0.2cm} 
\section{Detailed Analysis of the Fusion Performance}
\label{appendix C}

This appendix provides a detailed analysis of the performance metrics for all dual-modality and tri-modality fusion combinations evaluated in this study, extending the discussion presented in the paper.  This comprehensive analysis serves to justify the selection of the best-performing models and offers a broader understanding of the comparative effectiveness of different fusion strategies and modality combinations.

\begin{table*}[]
\centering
\small
\caption{Performance metrics for two-modal fusions under real and noisy augmented data.}
\vspace{-0.4cm} 
\label{tab:fusion_performance_two_modal_best_both}
\scalebox{0.9}{
\begin{tabular}{@{}llllcccccccccc@{}}
\toprule
\multicolumn{3}{c}{} & \multicolumn{5}{c}{\textbf{Real Data}} & \multicolumn{5}{c}{\textbf{Noisy Augmented Data}} & \multirow{2}{*}{\textbf{Detec. Time}} \\
\cmidrule(lr){4-8} \cmidrule(lr){9-13}
\textbf{Fusion} & \textbf{Modality} & \textbf{Models} & \textbf{Acc} & \textbf{Prec} & \textbf{Rec} & \textbf{F1} & \textbf{F1-M} & \textbf{Acc} & \textbf{Prec} & \textbf{Rec} & \textbf{F1} & \textbf{F1-M} & \multirow{2}{*}{\textbf{(ms)}} \\
\textbf{Type} & \textbf{Combination} & & \textbf{(\%)} & \textbf{(\%)} & \textbf{(\%)} & \textbf{(\%)} & \textbf{(\%)} & \textbf{(\%)} & \textbf{(\%)} & \textbf{(\%)} & \textbf{(\%)} & \textbf{(\%)} & \\
\midrule
\multirow{14}{*}{\rotatebox[origin=c]{90}{\textbf{Late}}} & \multirow{4}{*}{\textbf{Audio-Visual}} & LeNet + ResNet-10 & 96.97 & 95.45 & 100.00 & 97.67 & 96.66 & 36.36 & 0.00 & 0.00 & 0.00 & 26.67 & 3.26 \\
& & LeNet + MobileNet & 100.00 & 100.00 & 100.00 & 100.00 & 100.00 & 65.23 & 70.68 & 77.50 & 73.94 & 60.86 & 2.55 \\
& & VGG-19 + ResNet-10 & 96.97 & 95.45 & 100.00 & 97.67 & 96.66 & 66.82 & 66.05 & 98.45 & 79.06 & 49.57 & 5.22 \\
& & VGG-19 + MobileNet & 96.97 & 95.45 & 100.00 & 97.67 & 96.66 & 56.36 & 100.00 & 31.43 & 47.83 & 55.16 & 3.61 \\
\cmidrule(lr){2-14}
& \multirow{4}{*}{\textbf{Audio-RF}} & LeNet + ResNet-10 & 97.20 & 99.88 & 95.71 & 97.75 & 97.02 & 65.53 & 64.91 & 99.76 & 78.65 & 44.63 & 3.04 \\
& & LeNet + MobileNet & 98.03 & 99.39 & 97.50 & 98.44 & 97.89 & 63.64 & 63.64 & 100.00 & 77.78 & 38.89 & 3.40 \\
& & VGG-19 + ResNet-10 & 96.82 & 100.00 & 95.00 & 97.44 & 96.62 & 73.71 & 85.37 & 70.83 & 77.42 & 72.98 & 4.30 \\
& & VGG-19 + MobileNet & 96.67 & 99.38 & 95.36 & 97.33 & 96.45 & 63.64 & 63.64 & 100.00 & 77.78 & 38.89 & 3.58 \\
\cmidrule(lr){2-14}
& \multirow{4}{*}{\textbf{Visual-RF}} & ResNet-10 + ResNet-10 & 96.97 & 95.45 & 100.00 & 97.67 & 96.66 & 63.18 & 64.11 & 95.71 & 76.79 & 43.89 & 4.29 \\
& & ResNet-10 + MobileNet & 96.97 & 95.45 & 100.00 & 97.67 & 96.66 & 63.64 & 63.64 & 100.00 & 77.78 & 38.89 & 4.36 \\
& & MobileNet + ResNet-10 & 96.97 & 95.66 & 99.76 & 97.67 & 96.67 & 72.50 & 98.57 & 57.62 & 72.73 & 72.50 & 3.32 \\
& & MobileNet + MobileNet & 99.62 & 100.00 & 99.40 & 99.70 & 99.59 & 62.42 & 63.23 & 97.86 & 76.82 & 38.81 & 2.25 \\
\midrule
\multirow{14}{*}{\rotatebox[origin=c]{90}{\textbf{GMU}}} & \multirow{4}{*}{\textbf{Audio-Visual}} & LeNet + ResNet-10 & 89.77 & 97.96 & 85.71 & 91.43 & 89.38 & 63.64 & 63.64 & 100.00 & 77.78 & 38.89 & 3.36 \\
& & LeNet + MobileNet & 99.32 & 98.94 & 100.00 & 99.47 & 99.26 & 55.53 & 60.53 & 86.55 & 71.24 & 36.62 & 2.40 \\
& & VGG-19 + ResNet-10 & 96.97 & 95.45 & 100.00 & 97.67 & 96.66 & 42.35 & 78.01 & 13.10 & 22.43 & 38.28 & 4.53 \\
& & VGG-19 + MobileNet & 96.97 & 95.45 & 100.00 & 97.67 & 96.66 & 36.36 & 0.00 & 0.00 & 0.00 & 26.67 & 3.63 \\
\cmidrule(lr){2-14}
& \multirow{4}{*}{\textbf{Audio-RF}} & LeNet + ResNet-10 & 97.05 & 99.88 & 95.48 & 97.63 & 96.86 & 59.47 & 63.70 & 84.40 & 72.61 & 47.37 & 3.09 \\
& & LeNet + MobileNet & 97.05 & 99.51 & 95.83 & 97.63 & 96.85 & 63.64 & 63.64 & 100.00 & 77.78 & 38.89 & 2.37 \\
& & VGG-19 + ResNet-10 & 97.12 & 100.00 & 95.48 & 97.69 & 96.94 & 80.53 & 92.31 & 75.71 & 83.19 & 80.03 & 4.25 \\
& & VGG-19 + MobileNet & 94.17 & 95.36 & 95.48 & 95.42 & 93.70 & 61.89 & 63.62 & 93.69 & 75.78 & 43.22 & 3.43 \\
\cmidrule(lr){2-14}
& \multirow{4}{*}{\textbf{Visual-RF}} & ResNet-10 + ResNet-10 & 96.29 & 99.25 & 94.88 & 97.02 & 96.05 & 59.92 & 63.68 & 86.19 & 73.24 & 46.73 & 3.65 \\
& & ResNet-10 + MobileNet & 96.74 & 99.38 & 95.48 & 97.39 & 96.53 & 60.76 & 62.70 & 94.64 & 75.43 & 39.03 & 4.56 \\
& & MobileNet + ResNet-10 & 96.97 & 95.45 & 100.00 & 97.67 & 96.66 & 47.42 & 100.00 & 17.38 & 29.61 & 43.83 & 3.16 \\
& & MobileNet + MobileNet & 95.38 & 100.00 & 92.74 & 96.23 & 95.13 & 62.42 & 63.23 & 97.86 & 76.82 & 38.81 & 2.27 \\
\bottomrule
\end{tabular}
}
\vspace{-0.4cm} 
\end{table*}

\vspace{-0.31cm} 
\subsection{Dual-Modal Fusion}
\label{Appendix C1- Dual-Modality}
Table \ref{tab:fusion_performance_two_modal_best_both} presents the performance metrics for all evaluated dual-modality fusion models under both real and noisy augmented data, categorized by fusion type and modality combination.

\noindent \tikz[baseline=(char.base)]\node[shape=circle, fill=black, inner sep=1pt, text=white] (char) {1}; \textbf{Late Fusion.}~In real data, all dual-modality combinations demonstrate high accuracy, generally above 96\%, with audio-visual (LeNet + MobileNet) achieving a 100\% accuracy. This indicates that Late Fusion effectively leverages complementary information from different modalities when data quality is high. In noisy augmented conditions, however, performance varies significantly.

\noindent $\bullet$ \textbf{Audio-Visual.}~While achieving perfect accuracy in real data with LeNet + MobileNet, performance drops to 65.23\% in noisy augmented conditions. The VGG-19 + ResNet-10 combination shows slightly better noisy condition accuracy at 66.82\%, suggesting VGG-19 and ResNet-10 models are somewhat more robust to noise in a Late Fusion setup for audio-visual data, although still significantly degraded compared to real data. The other audio-visual combinations show even lower noisy condition performance, indicating that simply combining audio and visual modalities via Late Fusion does not inherently guarantee robustness in noisy environments.

\noindent $\bullet$ \textbf{Audio-RF.}~This modality combination generally exhibits more consistent performance across real and noisy augmneted data compared to audio-visual. The VGG-19 + ResNet-10 achieves the highest noisy condition accuracy among all dual-modality combinations, at 73.71\%. This suggests that combining audio and RF modalities in a Late Fusion manner provides a more robust approach to noise compared to audio-visual combinations. While LeNet + MobileNet achieves a slightly higher real data accuracy (98.03\%) compared to VGG-19 + ResNet-10 (96.82\%), its noisy condition performance (63.64\%) is considerably lower, further emphasizing the robustness of the VGG-19 + ResNet-10.

\noindent $\bullet$ \textbf{Visual-RF.}~Visual-RF combinations show strong performance in real data, with MobileNet + MobileNet reaching 99.62\% accuracy. However, in noisy conditions, their performance is comparable to or slightly lower than audio-visual. The MobileNet + ResNet-10 combination achieves the highest noisy condition accuracy among visual-RF at 72.50\%, which is still slightly below the best audio-RF (VGG-19 + ResNet-10 at 73.71\%). This indicates that while visual-RF fusion is effective, it does not offer superior robustness in noise compared to audio-RF fusion.

\begin{table*}[!ht]
\centering
\small
\caption{Performance metrics for three-modal fusions under real and noisy augmented data.}
\vspace{-0.4cm} 
\label{tab:fusion_performance_three_modal_best_both_avg_time_single_model_col}
\begin{adjustbox}{width=\textwidth,center}
\begin{tabular}{@{}llllcccccccccc@{}}
\toprule
\multicolumn{3}{c}{} & \multicolumn{5}{c}{\textbf{Real Data}} & \multicolumn{5}{c}{\textbf{Noisy Augmented Data}} & \multirow{2}{*}{\textbf{Detec. Time}} \\
\cmidrule(lr){4-8} \cmidrule(lr){9-13}
\textbf{Fusion} & \textbf{Modality} & \textbf{Models} & \textbf{Acc} & \textbf{Prec} & \textbf{Rec} & \textbf{F1} & \textbf{F1-M} & \textbf{Acc} & \textbf{Prec} & \textbf{Rec} & \textbf{F1} & \textbf{F1-M} & \multirow{2}{*}{\textbf{(ms)}} \\
\textbf{Type} & \textbf{Combination} & & \textbf{(\%)} & \textbf{(\%)} & \textbf{(\%)} & \textbf{(\%)} & \textbf{(\%)} & \textbf{(\%)} & \textbf{(\%)} & \textbf{(\%)} & \textbf{(\%)} & \textbf{(\%)} & \\
\midrule
\multirow{8}{*}{\rotatebox[origin=c]{90}{\textbf{Late}}} & \multirow{8}{*}{\textbf{Audio-Visual-RF}} & LeNet + ResNet-10 + ResNet-10 & 100.00 & 100.00 & 100.00 & 100.00 & 100.00 & 67.95 & 66.67 & 99.29 & 79.77 & 51.36 & 5.43 \\
& & LeNet + ResNet-10 + MobileNet & 100.00 & 100.00 & 100.00 & 100.00 & 100.00 & 63.18 & 63.49 & 99.17 & 77.42 & 38.91 & 7.89 \\
& & LeNet + MobileNet + ResNet-10 & 99.62 & 100.00 & 99.40 & 99.70 & 99.59 & 80.98 & 95.24 & 73.81 & 83.17 & 80.66 & 5.57 \\
& & LeNet + MobileNet + MobileNet & 100.00 & 100.00 & 100.00 & 100.00 & 100.00 & 65.00 & 64.65 & 99.29 & 78.31 & 43.86 & 3.70 \\
& & VGG-19 + ResNet-10 + ResNet-10 & 96.97 & 95.45 & 100.00 & 97.67 & 96.66 & 67.27 & 66.04 & 100.00 & 79.55 & 48.86 & 6.85 \\
& & VGG-19 + ResNet-10 + MobileNet & 95.61 & 93.54 & 100.00 & 96.66 & 95.12 & 57.35 & 61.67 & 87.14 & 72.22 & 40.19 & 8.85 \\
& & VGG-19 + MobileNet + ResNet-10 & 96.89 & 95.35 & 100.00 & 97.62 & 96.58 & 83.26 & 97.69 & 75.48 & 85.16 & 82.98 & 6.09 \\
& & VGG-19 + MobileNet + MobileNet & 100.00 & 100.00 & 100.00 & 100.00 & 100.00 & 63.64 & 63.64 & 100.00 & 77.78 & 38.89 & 4.64 \\
\midrule
\multirow{8}{*}{\rotatebox[origin=c]{90}{\textbf{GMU}}} & \multirow{8}{*}{\textbf{Audio-Visual-RF}} & LeNet + ResNet-10 + ResNet-10 & 96.97 & 95.45 & 100.00 & 97.67 & 96.66 & 75.53 & 72.23 & 100.00 & 83.87 & 66.58 & 6.48 \\
& & LeNet + ResNet-10 + MobileNet & 96.97 & 95.45 & 100.00 & 97.67 & 96.66 & 74.09 & 71.07 & 100.00 & 83.09 & 63.87 & 4.81 \\
& & LeNet + MobileNet + ResNet-10 & 97.20 & 99.51 & 96.07 & 97.76 & 97.01 & 68.48 & 69.78 & 89.05 & 78.24 & 60.55 & 4.95 \\
& & LeNet + MobileNet + MobileNet & 99.39 & 99.06 & 100.00 & 99.53 & 99.34 & 63.64 & 63.64 & 100.00 & 77.78 & 38.89 & 3.80 \\
& & VGG-19 + ResNet-10 + ResNet-10 & 97.05 & 95.56 & 100.00 & 97.73 & 96.75 & 63.64 & 63.64 & 100.00 & 77.78 & 38.89 & 7.55 \\
& & VGG-19 + ResNet-10 + MobileNet & 97.27 & 100.00 & 95.71 & 97.81 & 97.10 & 53.26 & 63.58 & 62.14 & 62.85 & 49.92 & 7.38 \\
& & VGG-19 + MobileNet + ResNet-10 & 96.97 & 95.45 & 100.00 & 97.67 & 96.66 & 36.67 & 100.00 & 0.48 & 0.01 & 27.20 & 5.67 \\
& & VGG-19 + MobileNet + MobileNet & 96.97 & 95.45 & 100.00 & 97.67 & 96.66 & 68.48 & 81.45 & 65.36 & 72.52 & 67.79 & 4.38 \\
\bottomrule
\end{tabular}
\end{adjustbox}
\vspace{-0.4cm} 
\end{table*}

\noindent \tikz[baseline=(char.base)]\node[shape=circle, fill=black, inner sep=1pt, text=white] (char) {2}; \textbf{GMU Fusion.}~In real data, GMU Fusion combinations also achieve high accuracy, although generally slightly lower than the best Late Fusion counterparts.  However, GMU Fusion demonstrates a different trend in noisy conditions.

\noindent $\bullet$ \textbf{Audio-Visual.}~In noisy augmented conditions, the audio-visual of GMU Fusion performance is generally lower than audio-visual in Late Fusion. The best-performing combination, LeNet + ResNet-10, achieves 63.64\% accuracy, significantly lower than the best audio-visual Late Fusion (66.82\%). This suggests that GMU Fusion may not be as effective as Late Fusion for audio-visual data in high noisy environments. Notably, the VGG-19 + ResNet-10 combination performs particularly poorly with GMU Fusion in noisy augmented conditions (42.35\% accuracy), indicating that complex models combined with GMU Fusion may not generalize well for audio-visual data under noise.

\noindent $\bullet$ \textbf{Audio-RF.}~Audio-RF fusion stands out as the most robust dual-modality approach in noisy conditions.  The VGG-19 + ResNet-10 combination achieves the highest noisy condition accuracy among all dual-modality combinations at 80.53\%. This significantly outperforms all Late Fusion dual-modality combinations in noise and demonstrates the effectiveness of GMU Fusion in leveraging audio and RF data for robust drone detection.  While LeNet + MobileNet achieves a lower noisy condition accuracy (63.64\%), it is still comparable to the most Late Fusion combinations.

\noindent $\bullet$ \textbf{Visual-RF.}~Visual-RF combinations show relatively consistent performance across real and noisy augmented data, but their noisy condition accuracy is generally lower than the best audio-RF GMU Fusion. The MobileNet + MobileNet combination, achieving 62.42\% accuracy in noisy conditions, is representative of the typical performance level for visual-RF GMU Fusion. While more robust than audio-visual, visual-RF does not reach the superior noisy condition performance of audio-RF.

\vspace{-0.35cm} 
\subsection{Tri-Modal Fusion}
\label{Appendix C2- Tri-Modality}
Table \ref{tab:fusion_performance_three_modal_best_both_avg_time_single_model_col} presents a detailed breakdown of the performance metrics for all evaluated tri-modality fusion models under both real and noisy augmented data. This table allows for a comprehensive comparison across different model combinations and fusion techniques when integrating audio, visual, and RF data.

\noindent \tikz[baseline=(char.base)]\node[shape=circle, fill=black, inner sep=1pt, text=white] (char) {1}; \textbf{Late Fusion.}~In real data, all tri-modality combinations achieve near-perfect accuracy, with several combinations reaching a perfect 100\% accuracy. This indicates that Late Fusion is highly effective in leveraging the synergistic strengths of audio, visual, and RF modalities when data quality is optimal. In noisy conditions, performance variation is observed, but overall, Late Fusion demonstrates significant robustness.

\noindent $\bullet$ \textbf{Audio-Visual-RF.}~The VGG-19 + MobileNet + ResNet-10 achieves the highest noisy condition accuracy at 83.26\%. This combination leverages the robust feature extraction capabilities of VGG-19 for audio, MobileNet for video, and ResNet-10 for RF, resulting in superior performance even when high noise is present. LeNet + MobileNet + ResNet-10 also performs strongly in noisy conditions with 80.98\% accuracy. Both combinations remain significantly above the best dual-modality noisy condition accuracy (80.53\% for audio-RF GMU Fusion), highlighting the added robustness of tri-modality Late Fusion. 
Notably, combinations including MobileNet sometimes tend to have lower detection times due to MobileNet's efficiency, with LeNet + MobileNet + MobileNet achieving the fastest detection time at 3.70 ms while still maintaining 65.00\% accuracy in noisy conditions.

\noindent \tikz[baseline=(char.base)]\node[shape=circle, fill=black, inner sep=1pt, text=white] (char) {2}; \textbf{GMU Fusion.}~In real data, tri-modality GMU Fusion combinations also exhibit high accuracy, comparable to Late Fusion. However, in noisy conditions, GMU Fusion shows a different performance profile.

\noindent $\bullet$ \textbf{Audio-Visual-RF.}~Among GMU Fusion combinations, LeNet + ResNet-10 + ResNet-10 achieves the highest noisy condition accuracy at 75.53\%. LeNet + ResNet-10 + MobileNet also performs well at 74.09\% accuracy in noisy conditions. While these accuracies are lower than the best Late Fusion tri-modality combination (83.26\%), they also remain below the best dual-modality performance under noisy conditions (80.53\%). This indicates that GMU Fusion, when applied to tri-modal data, is a weaker alternative to Late Fusion, exhibiting reduced peak accuracy in noisy environments. 

\end{document}

%% file: preamble.tex
\usepackage{amsmath,amsfonts}
\usepackage{algorithm}
\usepackage{algorithmic}
\usepackage{graphicx}
\usepackage{textcomp}
\usepackage{balance}
\usepackage{array}
\usepackage{enumitem}
\usepackage{subcaption}
\usepackage{tikz}
\usepackage{tikzscale}
\usepackage{pgfplots}
\usepackage{pgfplotstable}
\usepackage{adjustbox}
\usepackage{hyperref}
\usepackage{booktabs}
\usepackage{siunitx}
\usepackage{multirow}

\usepgfplotslibrary{colorbrewer,patchplots,groupplots}
\pgfplotsset{compat=1.18}
\newlength\fheight
\newlength\fwidth
\usetikzlibrary{plotmarks,patterns,decorations.pathreplacing,backgrounds,calc,arrows,arrows.meta,spy,matrix}

%% file: confusion_matrices/Scenario_2_Unimodal/lenet_audio.tex
\begin{tikzpicture}
\pgfplotsset{every tick label/.append style={font=\tiny}}

\begin{axis}[
    title={\textbf{Audio}},  
    title style={
      font=\scriptsize\color{white!10!black}\small,
      yshift=-1.2ex    
    },
    enlargelimits=false,
    colormap/PuBu,
    width=\fwidth,
    height=\fheight,
    scale only axis,
    tick align=inside,
    xlabel={Predicted Class},
    xlabel style={font=\scriptsize\color{white!10!black}\small},
    ylabel={Actual Class},
    ylabel style={font=\scriptsize\color{white!10!black}\small},
    xmin=0.5, xmax=2.5,
    ymin=0.5, ymax=2.5,
    xtick={1,2},
    ytick={1,2},
    xticklabels={0, 1},
    yticklabels={0, 1},
    xlabel shift=-5pt,
    ylabel shift=-5pt,
    ytick style={draw=none, line width=0.3pt},
    xtick style={line width=0.3pt},
    axis background/.style={fill=white},
    colorbar style={
      ticklabel style={font=\scriptsize},
      at={(0,1.05)},
      anchor=below south west,
      width=\pgfkeysvalueof{/pgfplots/parent axis width},
      xmin=0, xmax=100,
      xtick={0, 50, 100},
      xticklabels={0\%, 50\%, 100\%},
      axis x line*=top,
      point meta min=0,
      point meta max=100,
    },
    colorbar/width=2mm,
    xticklabel style={font=\scriptsize\color{white!15!black}\footnotesize},
    yticklabel style={font=\scriptsize\color{white!15!black}\footnotesize},
]

\addplot [
    matrix plot,
    point meta=explicit,
    nodes near coords={
        \pgfmathtruncatemacro{\x}{\pgfkeysvalueof{/data point/x}}%
        \pgfmathtruncatemacro{\y}{\pgfkeysvalueof{/data point/y}}%
        \ifnum\x=1
            \textcolor{white}{\pgfmathprintnumber{\pgfplotspointmeta}\%}%
        \else%
            \textcolor{black}{\pgfmathprintnumber{\pgfplotspointmeta}\%}%
        \fi%
    },
    nodes near coords align={center},
    every node near coord/.append style={
        font=\scriptsize,
        /pgf/number format/fixed,
        /pgf/number format/precision=1,
        /pgf/number format/fixed zerofill,
        /pgf/number format/1000 sep={}
    },
]
coordinates {
    (1,1) [100.0]  (2,1) [0.0]
    
    (1,2) [100.0]   (2,2) [0.0]

};

\end{axis}
\end{tikzpicture}

%% file: confusion_matrices/Scenario_2_Unimodal/resnet10_video.tex
\begin{tikzpicture}
\pgfplotsset{every tick label/.append style={font=\tiny}}

\begin{axis}[
    title={\textbf{Visual}},  
    title style={
      font=\scriptsize\color{white!10!black}\small,
      yshift=-1.2ex    
    },
    enlargelimits=false,
    colormap/YlGn,
    width=\fwidth,
    height=\fheight,
    scale only axis,
    tick align=inside,
    xlabel={Predicted Class},
    xlabel style={font=\scriptsize\color{white!10!black}\small},
    ylabel style={font=\scriptsize\color{white!10!black}\small},
    xmin=0.5, xmax=2.5,
    ymin=0.5, ymax=2.5,
    xtick={1,2},
    ytick={1,2},
    xticklabels={0, 1},
    yticklabels={0, 1},
    xlabel shift=-5pt,
    ylabel shift=-5pt,
    ytick style={draw=none, line width=0.3pt},
    xtick style={line width=0.3pt},
    axis background/.style={fill=white},
    colorbar style={
      ticklabel style={font=\scriptsize},
      at={(0,1.05)},
      anchor=below south west,
      width=\pgfkeysvalueof{/pgfplots/parent axis width},
      xmin=0, xmax=100,
      xtick={0, 50, 100},
      xticklabels={0\%, 50\%, 100\%},
      axis x line*=top,
      point meta min=0,
      point meta max=100,
    },
    colorbar/width=2mm,
    xticklabel style={font=\scriptsize\color{white!15!black}\footnotesize},
    yticklabel style={font=\scriptsize\color{white!15!black}\footnotesize},
]

\addplot [
    matrix plot,
    point meta=explicit,
    nodes near coords={
        \pgfmathtruncatemacro{\x}{\pgfkeysvalueof{/data point/x}}%
        \pgfmathtruncatemacro{\y}{\pgfkeysvalueof{/data point/y}}%
        \ifnum\x=1
            \textcolor{white}{\pgfmathprintnumber{\pgfplotspointmeta}\%}%
        \else%
            \textcolor{black}{\pgfmathprintnumber{\pgfplotspointmeta}\%}%
        \fi%
    },
    nodes near coords align={center},
    every node near coord/.append style={
        font=\scriptsize,
        /pgf/number format/fixed,
        /pgf/number format/precision=1,
        /pgf/number format/fixed zerofill,
        /pgf/number format/1000 sep={}
    },
]
coordinates {
    (1,1) [100.0]  (2,1) [0.0]
    
    (1,2) [98.95]   (2,2) [0.05]

};

\end{axis}
\end{tikzpicture}

%% file: confusion_matrices/Scenario_2_Unimodal/mobilenet_rf.tex
\begin{tikzpicture}
\pgfplotsset{every tick label/.append style={font=\tiny}}

\begin{axis}[
    title={\textbf{RF}},  
    title style={
      font=\scriptsize\color{white!10!black}\small,
      yshift=-1.2ex    
    },
    enlargelimits=false,
    colormap/Reds,
    width=\fwidth,
    height=\fheight,
    scale only axis,
    tick align=inside,
    xlabel={Predicted Class},
    xlabel style={font=\scriptsize\color{white!10!black}\small},
    ylabel style={font=\scriptsize\color{white!10!black}\small},
    xmin=0.5, xmax=2.5,
    ymin=0.5, ymax=2.5,
    xtick={1,2},
    ytick={1,2},
    xticklabels={0, 1},
    yticklabels={0, 1},
    xlabel shift=-5pt,
    ylabel shift=-5pt,
    ytick style={draw=none, line width=0.3pt},
    xtick style={line width=0.3pt},
    axis background/.style={fill=white},
    colorbar style={
      ticklabel style={font=\scriptsize},
      at={(0,1.05)},
      anchor=below south west,
      width=\pgfkeysvalueof{/pgfplots/parent axis width},
      xmin=0, xmax=100,
      xtick={0, 50, 100},
      xticklabels={0\%, 50\%, 100\%},
      axis x line*=top,
      point meta min=0,
      point meta max=100,
    },
    colorbar/width=2mm,
    xticklabel style={font=\scriptsize\color{white!15!black}\footnotesize},
    yticklabel style={font=\scriptsize\color{white!15!black}\footnotesize},
]

\addplot [
    matrix plot,
    point meta=explicit,
    nodes near coords={
        \pgfmathtruncatemacro{\x}{\pgfkeysvalueof{/data point/x}}%
        \pgfmathtruncatemacro{\y}{\pgfkeysvalueof{/data point/y}}%
        \ifnum\x=1
            \textcolor{white}{\pgfmathprintnumber{\pgfplotspointmeta}\%}%
        \else%
            \textcolor{black}{\pgfmathprintnumber{\pgfplotspointmeta}\%}%
        \fi%
    },
    nodes near coords align={center},
    every node near coord/.append style={
        font=\scriptsize,
        /pgf/number format/fixed,
        /pgf/number format/precision=1,
        /pgf/number format/fixed zerofill,
        /pgf/number format/1000 sep={}
    },
]
coordinates {
    (1,1) [92.14]  (2,1) [7.86]
    
    (1,2) [78.33]   (2,2) [21.7]

};

\end{axis}
\end{tikzpicture}

%% file: confusion_matrices/Dual_Modalities/Late/audio_visual_vgg_resnet10.tex
\begin{tikzpicture}
\pgfplotsset{every tick label/.append style={font=\tiny}}

\begin{axis}[
    title={\textbf{Audio-Visual}},  
    title style={
      font=\scriptsize\color{white!10!black}\small,
      yshift=-1.2ex    
    },
    enlargelimits=false,
    colormap/PuBu,
    width=\fwidth,
    height=\fheight,
    scale only axis,
    tick align=inside,
    xlabel={Predicted Class},
    xlabel style={font=\scriptsize\color{white!10!black}\small},
    ylabel={Actual Class},
    ylabel style={font=\scriptsize\color{white!10!black}\small},
    xmin=0.5, xmax=2.5,
    ymin=0.5, ymax=2.5,
    xtick={1,2},
    ytick={1,2},
    xticklabels={0, 1},
    yticklabels={0, 1},
    xlabel shift=-5pt,
    ylabel shift=-5pt,
    ytick style={draw=none, line width=0.3pt},
    xtick style={line width=0.3pt},
    axis background/.style={fill=white},
    colorbar style={
      ticklabel style={font=\scriptsize},
      at={(0,1.05)},
      anchor=below south west,
      width=\pgfkeysvalueof{/pgfplots/parent axis width},
      xmin=0, xmax=100,
      xtick={0, 50, 100},
      xticklabels={0\%, 50\%, 100\%},
      axis x line*=top,
      point meta min=0,
      point meta max=100,
    },
    colorbar/width=2mm,
    xticklabel style={font=\scriptsize\color{white!15!black}\footnotesize},
    yticklabel style={font=\scriptsize\color{white!15!black}\footnotesize},
]

\addplot [
    matrix plot,
    point meta=explicit,
    nodes near coords={
        \pgfmathtruncatemacro{\x}{\pgfkeysvalueof{/data point/x}}%
        \pgfmathtruncatemacro{\y}{\pgfkeysvalueof{/data point/y}}%
        \ifnum\x=1
            \textcolor{white}{\pgfmathprintnumber{\pgfplotspointmeta}\%}%
        \else%
            \textcolor{black}{\pgfmathprintnumber{\pgfplotspointmeta}\%}%
        \fi%
    },
    nodes near coords align={center},
    every node near coord/.append style={
        font=\scriptsize,
        /pgf/number format/fixed,
        /pgf/number format/precision=1,
        /pgf/number format/fixed zerofill,
        /pgf/number format/1000 sep={}
    },
]
coordinates {
    (1,1) [98.45]  (2,1) [1.5]
    
    (1,2) [88.54]   (2,2) [11.45]

};

\end{axis}
\end{tikzpicture}

%% file: confusion_matrices/Dual_Modalities/Gmu/audio_rf_vgg_resnet10.tex
\begin{tikzpicture}
\pgfplotsset{every tick label/.append style={font=\tiny}}

\begin{axis}[
    title={\textbf{Audio-RF}},  
    title style={
      font=\scriptsize\color{white!10!black}\small,
      yshift=-1.2ex    
    },
    enlargelimits=false,
    colormap/YlGn,
    width=\fwidth,
    height=\fheight,
    scale only axis,
    tick align=inside,
    xlabel={Predicted Class},
    xlabel style={font=\scriptsize\color{white!10!black}\small},
    ylabel style={font=\scriptsize\color{white!10!black}\small},
    xmin=0.5, xmax=2.5,
    ymin=0.5, ymax=2.5,
    xtick={1,2},
    ytick={1,2},
    xticklabels={0, 1},
    yticklabels={0, 1},
    xlabel shift=-5pt,
    ylabel shift=-5pt,
    ytick style={draw=none, line width=0.3pt},
    xtick style={line width=0.3pt},
    axis background/.style={fill=white},
    colorbar style={
      ticklabel style={font=\scriptsize},
      at={(0,1.05)},
      anchor=below south west,
      width=\pgfkeysvalueof{/pgfplots/parent axis width},
      xmin=0, xmax=100,
      xtick={0, 50, 100},
      xticklabels={0\%, 50\%, 100\%},
      axis x line*=top,
      point meta min=0,
      point meta max=100,
    },
    colorbar/width=2mm,
    xticklabel style={font=\scriptsize\color{white!15!black}\footnotesize},
    yticklabel style={font=\scriptsize\color{white!15!black}\footnotesize},
]

\addplot [
    matrix plot,
    point meta=explicit,
    nodes near coords={
        \pgfmathtruncatemacro{\x}{\pgfkeysvalueof{/data point/x}}%
        \pgfmathtruncatemacro{\y}{\pgfkeysvalueof{/data point/y}}%
        \ifnum\x=\y
            \textcolor{white}{\pgfmathprintnumber{\pgfplotspointmeta}\%}%
        \else%
            \textcolor{black}{\pgfmathprintnumber{\pgfplotspointmeta}\%}%
        \fi%
    },
    nodes near coords align={center},
    every node near coord/.append style={
        font=\scriptsize,
        /pgf/number format/fixed,
        /pgf/number format/precision=1,
        /pgf/number format/fixed zerofill,
        /pgf/number format/1000 sep={}
    },
]
coordinates {
    (1,1) [75.72]  (2,1) [24.29]
    
    (1,2) [11.04]   (2,2) [88.96]

};

\end{axis}
\end{tikzpicture}

%% file: confusion_matrices/Dual_Modalities/Late/visual_rf_mobilenet_resnet10.tex
\begin{tikzpicture}
\pgfplotsset{every tick label/.append style={font=\tiny}}

\begin{axis}[
    title={\textbf{Visual-RF}},  
    title style={
      font=\scriptsize\color{white!10!black}\small,
      yshift=-1.2ex    
    },
    enlargelimits=false,
    colormap/PuBu,
    width=\fwidth,
    height=\fheight,
    scale only axis,
    tick align=inside,
    xlabel={Predicted Class},
    xlabel style={font=\scriptsize\color{white!10!black}\small},
    ylabel style={font=\scriptsize\color{white!10!black}\small},
    xmin=0.5, xmax=2.5,
    ymin=0.5, ymax=2.5,
    xtick={1,2},
    ytick={1,2},
    xticklabels={0, 1},
    yticklabels={0, 1},
    xlabel shift=-5pt,
    ylabel shift=-5pt,
    ytick style={draw=none, line width=0.3pt},
    xtick style={line width=0.3pt},
    axis background/.style={fill=white},
    colorbar style={
      ticklabel style={font=\scriptsize},
      at={(0,1.05)},
      anchor=below south west,
      width=\pgfkeysvalueof{/pgfplots/parent axis width},
      xmin=0, xmax=100,
      xtick={0, 50, 100},
      xticklabels={0\%, 50\%, 100\%},
      axis x line*=top,
      point meta min=0,
      point meta max=100,
    },
    colorbar/width=2mm,
    xticklabel style={font=\scriptsize\color{white!15!black}\footnotesize},
    yticklabel style={font=\scriptsize\color{white!15!black}\footnotesize},
]

\addplot [
    matrix plot,
    point meta=explicit,
    nodes near coords={
        \pgfmathtruncatemacro{\x}{\pgfkeysvalueof{/data point/x}}%
        \pgfmathtruncatemacro{\y}{\pgfkeysvalueof{/data point/y}}%
        \ifnum\x=\y
            \textcolor{white}{\pgfmathprintnumber{\pgfplotspointmeta}\%}%
        \else%
            \textcolor{black}{\pgfmathprintnumber{\pgfplotspointmeta}\%}%
        \fi%
    },
    nodes near coords align={center},
    every node near coord/.append style={
        font=\scriptsize,
        /pgf/number format/fixed,
        /pgf/number format/precision=1,
        /pgf/number format/fixed zerofill,
        /pgf/number format/1000 sep={}
    },
]
coordinates {
    (1,1) [57.62]  (2,1) [42.38]
    
    (1,2) [1.46]   (2,2) [98.54]

};

\end{axis}
\end{tikzpicture}

%% file: confusion_matrices/late_3_modal/lenet_mobilenet_resnet10.tex
\begin{tikzpicture}
\pgfplotsset{every tick label/.append style={font=\tiny}}

\begin{axis}[
    enlargelimits=false,
    colormap/PuBu,
    width=\fwidth,
    height=\fheight,
    scale only axis,
    tick align=inside,
    xlabel={Predicted Class},
    xlabel style={font=\scriptsize\color{white!10!black}\small},
    ylabel={Actual Class},
    ylabel style={font=\scriptsize\color{white!10!black}\small},
    xmin=0.5, xmax=2.5,
    ymin=0.5, ymax=2.5,
    xtick={1,2},
    ytick={1,2},
    xticklabels={0, 1},
    yticklabels={0, 1},
    xlabel shift=-5pt,
    ylabel shift=-5pt,
    ytick style={draw=none, line width=0.3pt},
    xtick style={line width=0.3pt},
    axis background/.style={fill=white},
    colorbar style={
      ticklabel style={font=\scriptsize},
      at={(0,1.05)},
      anchor=below south west,
      width=\pgfkeysvalueof{/pgfplots/parent axis width},
      xmin=0, xmax=100,
      xtick={0, 50, 100},
      xticklabels={0\%, 50\%, 100\%},
      axis x line*=top,
      point meta min=0,
      point meta max=100,
    },
    colorbar/width=2mm,
    xticklabel style={font=\scriptsize\color{white!15!black}\footnotesize},
    yticklabel style={font=\scriptsize\color{white!15!black}\footnotesize},
]

\addplot [
    matrix plot,
    point meta=explicit,
    nodes near coords={
        \pgfmathtruncatemacro{\x}{\pgfkeysvalueof{/data point/x}}%
        \pgfmathtruncatemacro{\y}{\pgfkeysvalueof{/data point/y}}%
        \ifnum\x=\y%
            \textcolor{white}{\pgfmathprintnumber{\pgfplotspointmeta}\%}%
        \else%
            \textcolor{black}{\pgfmathprintnumber{\pgfplotspointmeta}\%}%
        \fi%
    },
    nodes near coords align={center},
    every node near coord/.append style={
        font=\scriptsize,
        /pgf/number format/fixed,
        /pgf/number format/precision=1,
        /pgf/number format/fixed zerofill,
        /pgf/number format/1000 sep={}
    },
]
coordinates {
    (1,1) [75.48]  (2,1) [24.52]
    
    (1,2) [3.13]   (2,2) [96.88]

};

\end{axis}
\end{tikzpicture}

%% file: confusion_matrices/gmu_3_modal/lenet_resnet10_resnet10.tex
\begin{tikzpicture}
\pgfplotsset{every tick label/.append style={font=\tiny}}

\begin{axis}[
    enlargelimits=false,
    colormap/YlGn ,
    width=\fwidth,
    height=\fheight,
    scale only axis,
    tick align=inside,
    xlabel={Predicted Class},
    xlabel style={font=\scriptsize\color{white!10!black}\small},
    ylabel style={font=\scriptsize\color{white!10!black}\small},
    xmin=0.5, xmax=2.5,
    ymin=0.5, ymax=2.5,
    xtick={1,2},
    ytick={1,2},
    xticklabels={0, 1},
    yticklabels={0, 1},
    xlabel shift=-5pt,
    ylabel shift=-5pt,
    ytick style={draw=none, line width=0.3pt},
    xtick style={line width=0.3pt},
    axis background/.style={fill=white},
    colorbar style={
      ticklabel style={font=\scriptsize},
      at={(0,1.05)},
      anchor=below south west,
      width=\pgfkeysvalueof{/pgfplots/parent axis width},
      xmin=0, xmax=100,
      xtick={0, 50, 100},
      xticklabels={0\%, 50\%, 100\%},
      axis x line*=top,
      point meta min=0,
      point meta max=100,
    },
    colorbar/width=2mm,
    xticklabel style={font=\scriptsize\color{white!15!black}\footnotesize},
    yticklabel style={font=\scriptsize\color{white!15!black}\footnotesize},
]

\addplot [
    matrix plot,
    point meta=explicit,
    nodes near coords={
        \pgfmathtruncatemacro{\x}{\pgfkeysvalueof{/data point/x}}%
        \ifnum\x=1
            \textcolor{white}{\pgfmathprintnumber{\pgfplotspointmeta}\%}%
        \else
            \textcolor{black}{\pgfmathprintnumber{\pgfplotspointmeta}\%}%
        \fi
    },
    nodes near coords align={center},
    every node near coord/.append style={
        font=\scriptsize,
        /pgf/number format/fixed,
        /pgf/number format/precision=1,
        /pgf/number format/fixed zerofill,
        /pgf/number format/1000 sep={}
    },
]
coordinates {
    (1,1) [100.0]  (2,1) [0.0]
    
    (1,2) [67.3]   (2,2) [32.7]

};

\end{axis}
\end{tikzpicture}

%% file: confusion_matrices/Day/late_VGG_+_MobileNet_+_ResNet10.tex
\begin{tikzpicture}
\pgfplotsset{every tick label/.append style={font=\tiny}}

\begin{axis}[
    enlargelimits=false,
    colormap/PuBu,
    width=\fwidth,
    height=\fheight,
    scale only axis,
    tick align=inside,
    xlabel={Predicted Class},
    xlabel style={font=\scriptsize\color{white!10!black}\small},
    ylabel={Actual Class},
    ylabel style={font=\scriptsize\color{white!10!black}\small},
    xmin=0.5, xmax=2.5,
    ymin=0.5, ymax=2.5,
    xtick={1,2},
    ytick={1,2},
    xticklabels={0, 1},
    yticklabels={0, 1},
    xlabel shift=-5pt,
    ylabel shift=-5pt,
    ytick style={draw=none, line width=0.3pt},
    xtick style={line width=0.3pt},
    axis background/.style={fill=white},
    colorbar style={
      ticklabel style={font=\scriptsize},
      at={(0,1.05)},
      anchor=below south west,
      width=\pgfkeysvalueof{/pgfplots/parent axis width},
      xmin=0, xmax=100,
      xtick={0, 50, 100},
      xticklabels={0\%, 50\%, 100\%},
      axis x line*=top,
      point meta min=0,
      point meta max=100,
    },
    colorbar/width=2mm,
    xticklabel style={font=\scriptsize\color{white!15!black}\footnotesize},
    yticklabel style={font=\scriptsize\color{white!15!black}\footnotesize},
]

\addplot [
    matrix plot,
    point meta=explicit,
    nodes near coords={
        \pgfmathtruncatemacro{\x}{\pgfkeysvalueof{/data point/x}}%
        \pgfmathtruncatemacro{\y}{\pgfkeysvalueof{/data point/y}}%
        \ifnum\x=\y%
            \textcolor{white}{\pgfmathprintnumber{\pgfplotspointmeta}\%}%
        \else%
            \textcolor{black}{\pgfmathprintnumber{\pgfplotspointmeta}\%}%
        \fi%
    },
    nodes near coords align={center},
    every node near coord/.append style={
        font=\scriptsize,
        /pgf/number format/fixed,
        /pgf/number format/precision=1,
        /pgf/number format/fixed zerofill,
        /pgf/number format/1000 sep={}
    },
]
coordinates {
    (1,1) [57.22]   (2,1) [42.78]

    (1,2) [0.625]   (2,2) [99.375]
};

\end{axis}
\end{tikzpicture}

%% file: confusion_matrices/Sunset/Late_VGG_+_MobileNet_+_ResNet10.tex
\begin{tikzpicture}
\pgfplotsset{every tick label/.append style={font=\tiny}}

\begin{axis}[
    enlargelimits=false,
    colormap/PuBu,
    width=\fwidth,
    height=\fheight,
    scale only axis,
    tick align=inside,
    xlabel={Predicted Class},
    xlabel style={font=\scriptsize\color{white!10!black}\small},
    ylabel style={font=\scriptsize\color{white!10!black}\small},
    xmin=0.5, xmax=2.5,
    ymin=0.5, ymax=2.5,
    xtick={1,2},
    ytick={1,2},
    xticklabels={0, 1},
    yticklabels={0, 1},
    xlabel shift=-5pt,
    ylabel shift=-5pt,
    ytick style={draw=none, line width=0.3pt},
    xtick style={line width=0.3pt},
    axis background/.style={fill=white},
    colorbar style={
      ticklabel style={font=\scriptsize},
      at={(0,1.05)},
      anchor=below south west,
      width=\pgfkeysvalueof{/pgfplots/parent axis width},
      xmin=0, xmax=100,
      xtick={0, 50, 100},
      xticklabels={0\%, 50\%, 100\%},
      axis x line*=top,
      point meta min=0,
      point meta max=100,
    },
    colorbar/width=2mm,
    xticklabel style={font=\scriptsize\color{white!15!black}\footnotesize},
    yticklabel style={font=\scriptsize\color{white!15!black}\footnotesize},
]

\addplot [
    matrix plot,
    point meta=explicit,
    nodes near coords={
        \pgfmathtruncatemacro{\x}{\pgfkeysvalueof{/data point/x}}%
        \pgfmathtruncatemacro{\y}{\pgfkeysvalueof{/data point/y}}%
        \ifnum\x=\y%
            \textcolor{white}{\pgfmathprintnumber{\pgfplotspointmeta}\%}%
        \else%
            \textcolor{black}{\pgfmathprintnumber{\pgfplotspointmeta}\%}%
        \fi%
    },
    nodes near coords align={center},
    every node near coord/.append style={
        font=\scriptsize,
        /pgf/number format/fixed,
        /pgf/number format/precision=1,
        /pgf/number format/fixed zerofill,
        /pgf/number format/1000 sep={}
    },
]
coordinates {
    (1,1) [89.17]   (2,1) [10.83]

    (1,2) [8.125]   (2,2) [91.875]
};

\end{axis}
\end{tikzpicture}

%% file: confusion_matrices/Day/gmu_lenet_+_Resnet10_+_ResNet10.tex
\begin{tikzpicture}
\pgfplotsset{every tick label/.append style={font=\tiny}}

\begin{axis}[
    enlargelimits=false,
    colormap/YlGn,
    width=\fwidth,
    height=\fheight,
    scale only axis,
    tick align=inside,
    xlabel={Predicted Class},
    xlabel style={font=\scriptsize\color{white!10!black}\small},
    ylabel style={font=\scriptsize\color{white!10!black}\small},
    xmin=0.5, xmax=2.5,
    ymin=0.5, ymax=2.5,
    xtick={1,2},
    ytick={1,2},
    xticklabels={0, 1},
    yticklabels={0, 1},
    xlabel shift=-5pt,
    ylabel shift=-5pt,
    ytick style={draw=none, line width=0.3pt},
    xtick style={line width=0.3pt},
    axis background/.style={fill=white},
    colorbar style={
      ticklabel style={font=\scriptsize},
      at={(0,1.05)},
      anchor=below south west,
      width=\pgfkeysvalueof{/pgfplots/parent axis width},
      xmin=0, xmax=100,
      xtick={0, 50, 100},
      xticklabels={0\%, 50\%, 100\%},
      axis x line*=top,
      point meta min=0,
      point meta max=100,
    },
    colorbar/width=2mm,
    xticklabel style={font=\scriptsize\color{white!15!black}\footnotesize},
    yticklabel style={font=\scriptsize\color{white!15!black}\footnotesize},
]

\addplot [
    matrix plot,
    point meta=explicit,
    nodes near coords={
        \pgfmathtruncatemacro{\x}{\pgfkeysvalueof{/data point/x}}%
        \pgfmathtruncatemacro{\y}{\pgfkeysvalueof{/data point/y}}%
        \ifnum\x=1%
            \textcolor{white}{\pgfmathprintnumber{\pgfplotspointmeta}\%}%
        \else%
            \textcolor{black}{\pgfmathprintnumber{\pgfplotspointmeta}\%}%
        \fi%
    },
    nodes near coords align={center},
    every node near coord/.append style={
        font=\scriptsize,
        /pgf/number format/fixed,
        /pgf/number format/precision=1,
        /pgf/number format/fixed zerofill,
        /pgf/number format/1000 sep={}
    },
]
coordinates {
    (1,1) [99.72]   (2,1) [0.277]

    (1,2) [54.375]   (2,2) [45.625]
};

\end{axis}
\end{tikzpicture}

%% file: confusion_matrices/Sunset/GMU_LeNet_+_ResNet10_+_ResNet10.tex
\begin{tikzpicture}
\pgfplotsset{every tick label/.append style={font=\tiny}}

\begin{axis}[
    enlargelimits=false,
    colormap/YlGn,
    width=\fwidth,
    height=\fheight,
    scale only axis,
    tick align=inside,
    xlabel={Predicted Class},
    xlabel style={font=\scriptsize\color{white!10!black}\small},
    ylabel style={font=\scriptsize\color{white!10!black}\small},
    xmin=0.5, xmax=2.5,
    ymin=0.5, ymax=2.5,
    xtick={1,2},
    ytick={1,2},
    xticklabels={0, 1},
    yticklabels={0, 1},
    xlabel shift=-5pt,
    ylabel shift=-5pt,
    ytick style={draw=none, line width=0.3pt},
    xtick style={line width=0.3pt},
    axis background/.style={fill=white},
    colorbar style={
      ticklabel style={font=\scriptsize},
      at={(0,1.05)},
      anchor=below south west,
      width=\pgfkeysvalueof{/pgfplots/parent axis width},
      xmin=0, xmax=100,
      xtick={0, 50, 100},
      xticklabels={0\%, 50\%, 100\%},
      axis x line*=top,
      point meta min=0,
      point meta max=100,
    },
    colorbar/width=2mm,
    xticklabel style={font=\scriptsize\color{white!15!black}\footnotesize},
    yticklabel style={font=\scriptsize\color{white!15!black}\footnotesize},
]

\addplot [
    matrix plot,
    point meta=explicit,
    nodes near coords={
        \pgfmathtruncatemacro{\x}{\pgfkeysvalueof{/data point/x}}%
        \pgfmathtruncatemacro{\y}{\pgfkeysvalueof{/data point/y}}%
        \ifnum\x=1%
            \textcolor{white}{\pgfmathprintnumber{\pgfplotspointmeta}\%}%
        \else%
            \textcolor{black}{\pgfmathprintnumber{\pgfplotspointmeta}\%}%
        \fi%
    },
    nodes near coords align={center},
    every node near coord/.append style={
        font=\scriptsize,
        /pgf/number format/fixed,
        /pgf/number format/precision=1,
        /pgf/number format/fixed zerofill,
        /pgf/number format/1000 sep={}
    },
]
coordinates {
    (1,1) [100.00]   (2,1) [0.0]

    (1,2) [98.125]   (2,2) [1.875]
};

\end{axis}
\end{tikzpicture}

%% file: confusion_matrices/Urban/late_VGG_+_MobileNet_+_ResNet10.tex
\begin{tikzpicture}
\pgfplotsset{every tick label/.append style={font=\tiny}}

\begin{axis}[
    enlargelimits=false,
    colormap/PuBu,
    width=\fwidth,
    height=\fheight,
    scale only axis,
    tick align=inside,
    xlabel={Predicted Class},
    xlabel style={font=\scriptsize\color{white!10!black}\small},
    ylabel={Actual Class},
    ylabel style={font=\scriptsize\color{white!10!black}\small},
    xmin=0.5, xmax=2.5,
    ymin=0.5, ymax=2.5,
    xtick={1,2},
    ytick={1,2},
    xticklabels={0, 1},
    yticklabels={0, 1},
    xlabel shift=-5pt,
    ylabel shift=-5pt,
    ytick style={draw=none, line width=0.3pt},
    xtick style={line width=0.3pt},
    axis background/.style={fill=white},
    colorbar style={
      ticklabel style={font=\scriptsize},
      at={(0,1.05)},
      anchor=below south west,
      width=\pgfkeysvalueof{/pgfplots/parent axis width},
      xmin=0, xmax=100,
      xtick={0, 50, 100},
      xticklabels={0\%, 50\%, 100\%},
      axis x line*=top,
      point meta min=0,
      point meta max=100,
    },
    colorbar/width=2mm,
    xticklabel style={font=\scriptsize\color{white!15!black}\footnotesize},
    yticklabel style={font=\scriptsize\color{white!15!black}\footnotesize},
]

\addplot [
    matrix plot,
    point meta=explicit,
    nodes near coords={
        \pgfmathtruncatemacro{\x}{\pgfkeysvalueof{/data point/x}}%
        \pgfmathtruncatemacro{\y}{\pgfkeysvalueof{/data point/y}}%
        \ifnum\x=\y%
            \textcolor{white}{\pgfmathprintnumber{\pgfplotspointmeta}\%}%
        \else%
            \textcolor{black}{\pgfmathprintnumber{\pgfplotspointmeta}\%}%
        \fi%
    },
    nodes near coords align={center},
    every node near coord/.append style={
        font=\scriptsize,
        /pgf/number format/fixed,
        /pgf/number format/precision=1,
        /pgf/number format/fixed zerofill,
        /pgf/number format/1000 sep={}
    },
]
coordinates {
    (1,1) [61.25]   (2,1) [38.75]

    (1,2) [7.0]   (2,2) [93.0]
};

\end{axis}
\end{tikzpicture}

%% file: confusion_matrices/Non_Urban/late_VGG_+_MobileNet_+_ResNet10.tex
\begin{tikzpicture}
\pgfplotsset{every tick label/.append style={font=\tiny}}

\begin{axis}[
    enlargelimits=false,
    colormap/PuBu,
    width=\fwidth,
    height=\fheight,
    scale only axis,
    tick align=inside,
    xlabel={Predicted Class},
    xlabel style={font=\scriptsize\color{white!10!black}\small},
    ylabel style={font=\scriptsize\color{white!10!black}\small},
    xmin=0.5, xmax=2.5,
    ymin=0.5, ymax=2.5,
    xtick={1,2},
    ytick={1,2},
    xticklabels={0, 1},
    yticklabels={0, 1},
    xlabel shift=-5pt,
    ylabel shift=-5pt,
    ytick style={draw=none, line width=0.3pt},
    xtick style={line width=0.3pt},
    axis background/.style={fill=white},
    colorbar style={
      ticklabel style={font=\scriptsize},
      at={(0,1.05)},
      anchor=below south west,
      width=\pgfkeysvalueof{/pgfplots/parent axis width},
      xmin=0, xmax=100,
      xtick={0, 50, 100},
      xticklabels={0\%, 50\%, 100\%},
      axis x line*=top,
      point meta min=0,
      point meta max=100,
    },
    colorbar/width=2mm,
    xticklabel style={font=\scriptsize\color{white!15!black}\footnotesize},
    yticklabel style={font=\scriptsize\color{white!15!black}\footnotesize},
]

\addplot [
    matrix plot,
    point meta=explicit,
    nodes near coords={
        \pgfmathtruncatemacro{\x}{\pgfkeysvalueof{/data point/x}}%
        \pgfmathtruncatemacro{\y}{\pgfkeysvalueof{/data point/y}}%
        \ifnum\x=\y%
            \textcolor{white}{\pgfmathprintnumber{\pgfplotspointmeta}\%}%
        \else%
            \textcolor{black}{\pgfmathprintnumber{\pgfplotspointmeta}\%}%
        \fi%
    },
    nodes near coords align={center},
    every node near coord/.append style={
        font=\scriptsize,
        /pgf/number format/fixed,
        /pgf/number format/precision=1,
        /pgf/number format/fixed zerofill,
        /pgf/number format/1000 sep={}
    },
]
coordinates {
    (1,1) [78.82]   (2,1) [21.18]

    (1,2) [0.36]   (2,2) [99.64]
};

\end{axis}
\end{tikzpicture}

%% file: confusion_matrices/Urban/gmu_lenet_+_Resnet10_+_ResNet10.tex
\begin{tikzpicture}
\pgfplotsset{every tick label/.append style={font=\tiny}}

\begin{axis}[
    enlargelimits=false,
    colormap/YlGn,
    width=\fwidth,
    height=\fheight,
    scale only axis,
    tick align=inside,
    xlabel={Predicted Class},
    xlabel style={font=\scriptsize\color{white!10!black}\small},
    ylabel style={font=\scriptsize\color{white!10!black}\small},
    xmin=0.5, xmax=2.5,
    ymin=0.5, ymax=2.5,
    xtick={1,2},
    ytick={1,2},
    xticklabels={0, 1},
    yticklabels={0, 1},
    xlabel shift=-5pt,
    ylabel shift=-5pt,
    ytick style={draw=none, line width=0.3pt},
    xtick style={line width=0.3pt},
    axis background/.style={fill=white},
    colorbar style={
      ticklabel style={font=\scriptsize},
      at={(0,1.05)},
      anchor=below south west,
      width=\pgfkeysvalueof{/pgfplots/parent axis width},
      xmin=0, xmax=100,
      xtick={0, 50, 100},
      xticklabels={0\%, 50\%, 100\%},
      axis x line*=top,
      point meta min=0,
      point meta max=100,
    },
    colorbar/width=2mm,
    xticklabel style={font=\scriptsize\color{white!15!black}\footnotesize},
    yticklabel style={font=\scriptsize\color{white!15!black}\footnotesize},
]

\addplot [
    matrix plot,
    point meta=explicit,
    nodes near coords={
        \pgfmathtruncatemacro{\x}{\pgfkeysvalueof{/data point/x}}%
        \pgfmathtruncatemacro{\y}{\pgfkeysvalueof{/data point/y}}%
        \ifnum\x=1%
            \textcolor{white}{\pgfmathprintnumber{\pgfplotspointmeta}\%}%
        \else%
            \textcolor{black}{\pgfmathprintnumber{\pgfplotspointmeta}\%}%
        \fi%
    },
    nodes near coords align={center},
    every node near coord/.append style={
        font=\scriptsize,
        /pgf/number format/fixed,
        /pgf/number format/precision=1,
        /pgf/number format/fixed zerofill,
        /pgf/number format/1000 sep={}
    },
]
coordinates {
    (1,1) [100.0]   (2,1) [0.0]

    (1,2) [96.5]   (2,2) [3.5]
};

\end{axis}
\end{tikzpicture}

%% file: confusion_matrices/Non_Urban/gmu_lenet_+_Resnet10_+_ResNet10.tex
\begin{tikzpicture}
\pgfplotsset{every tick label/.append style={font=\tiny}}

\begin{axis}[
    enlargelimits=false,
    colormap/YlGn,
    width=\fwidth,
    height=\fheight,
    scale only axis,
    tick align=inside,
    xlabel={Predicted Class},
    xlabel style={font=\scriptsize\color{white!10!black}\small},
    ylabel style={font=\scriptsize\color{white!10!black}\small},
    xmin=0.5, xmax=2.5,
    ymin=0.5, ymax=2.5,
    xtick={1,2},
    ytick={1,2},
    xticklabels={0, 1},
    yticklabels={0, 1},
    xlabel shift=-5pt,
    ylabel shift=-5pt,
    ytick style={draw=none, line width=0.3pt},
    xtick style={line width=0.3pt},
    axis background/.style={fill=white},
    colorbar style={
      ticklabel style={font=\scriptsize},
      at={(0,1.05)},
      anchor=below south west,
      width=\pgfkeysvalueof{/pgfplots/parent axis width},
      xmin=0, xmax=100,
      xtick={0, 50, 100},
      xticklabels={0\%, 50\%, 100\%},
      axis x line*=top,
      point meta min=0,
      point meta max=100,
    },
    colorbar/width=2mm,
    xticklabel style={font=\scriptsize\color{white!15!black}\footnotesize},
    yticklabel style={font=\scriptsize\color{white!15!black}\footnotesize},
]

\addplot [
    matrix plot,
    point meta=explicit,
    nodes near coords={
        \pgfmathtruncatemacro{\x}{\pgfkeysvalueof{/data point/x}}%
        \pgfmathtruncatemacro{\y}{\pgfkeysvalueof{/data point/y}}%
        \ifnum\x=\y%
            \textcolor{white}{\pgfmathprintnumber{\pgfplotspointmeta}\%}%
        \else%
            \textcolor{black}{\pgfmathprintnumber{\pgfplotspointmeta}\%}%
        \fi%
    },
    nodes near coords align={center},
    every node near coord/.append style={
        font=\scriptsize,
        /pgf/number format/fixed,
        /pgf/number format/precision=1,
        /pgf/number format/fixed zerofill,
        /pgf/number format/1000 sep={}
    },
]
coordinates {
    (1,1) [100.0]   (2,1) [0.0]

    (1,2) [46.79]   (2,2) [53.21]
};

\end{axis}
\end{tikzpicture}